\documentclass[10pt,twocolumn,english,prl,aps,floatfix,superscriptaddress]{revtex4}
\usepackage{babel}
\usepackage{verbatim}
\usepackage{float}
\usepackage{amsmath}
\usepackage{amssymb}
\usepackage{graphicx}
\usepackage{color}

\usepackage[T1]{fontenc}
\usepackage{bbold}

\makeatletter
\@ifundefined{textcolor}{}
{%
 \definecolor{BLACK}{gray}{0}
 \definecolor{WHITE}{gray}{1}
 \definecolor{RED}{rgb}{1,0,0}
 \definecolor{GREEN}{rgb}{0,1,0}
 \definecolor{BLUE}{rgb}{0,0,1}
 \definecolor{CYAN}{cmyk}{1,0,0,0}
 \definecolor{MAGENTA}{cmyk}{0,1,0,0}
 \definecolor{YELLOW}{cmyk}{0,0,1,0}
}

\newcommand{\OM}{optomechanical }        
\newcommand{\SSH}{Su-Schrieffer-Heeger}
\newcommand{\LatJ}{J }        
\newcommand{\LatK}{K }       
\newcommand{\SSHJ}{K  }        
\newcommand{\SSHJP}{K_p }   
\newcommand{\blochHk}{h_{k} }
\newcommand{\params}{\Delta=-\Omega, g= 0.02 \Omega ,\kappa = 0.01 \Omega, \Gamma = 0.001 \Omega}
\newcommand{\NAdk}{k_{c}}
\newcommand{\gs}{G}
\newcommand{\ex}{E}
\newcommand{\oprt}[1]{\hat{#1}}

\newcommand{\Oprtk}[2]{\hat{#1}_{#2}}
\newcommand{\Oprtkg}[3]{\hat{#1}_{#2,#3}}
\newcommand{\Oprtkgt}[4]{\hat{#1}_{#2,#3} \left( #4\right)}

\newcommand{\ket}[1]{\left\vert #1\right\rangle}
\newcommand{\bra}[1]{\left\langle #1\right\vert}

\makeatother

\begin{document}

\title{Quench Dynamics in 1D Optomechanical Arrays}

\author{Sadegh Raeisi}

\email[]{sadegh.raeisi@gmail.com}

\affiliation{Department of Physics, Sharif University of Technology, Tehran, Iran }

\author{Florian Marquardt}

\affiliation{Max Planck Institute for the Science of Light, Staudtstra{\ss}e 2, 91058 Erlangen, Germany}

\affiliation{Institute for Theoretical Physics II, Friedrich-Alexander-Universit{\"a}t
Erlangen-N{\"u}rnberg, Staudtstra{\ss}e 7, 91058 Erlangen, Germany}

\begin{abstract}
Non-equilibrium dynamics induced by rapid 
changes of external parameters is relevant 
for a wide range of scenarios across many 
domains of physics. For waves in spatially 
periodic systems, quenches will alter the 
bandstructure and generate new excitations. 
In the case of topological bandstructures, 
defect modes at boundaries can be generated 
or destroyed when quenching through a 
topological phase transition. Here, we 
demonstrate that optomechanical arrays are 
a promising platform for studying such dynamics, 
as their bandstructure can be tuned temporally 
by a control laser. We study the creation of 
nonequilibrium optical and mechanical excitations 
in 1D arrays, including a bosonic version of the 
Su-Schrieffer-Heeger model. These ideas can 
be transferred to other systems such as 
driven nonlinear cavity arrays.
\end{abstract}
\maketitle

\section{Introduction}

Cavity optomechanics \cite{aspelmeyer2014cavity}
exploits the
radiation pressure interaction to couple 
optical and mechanical
degrees of freedom. A centerpiece of the 
physics encountered in
this setting is the parametric nature of 
the optomechanical interaction: the
radiation force is quadratic in the light 
amplitude. Upon driving
such a system by a control laser field, 
this results in an
effective laser-enhanced linear coupling 
between optics and mechanics. Importantly,
that coupling is tuneable by the control 
laser amplitude. This tuneability
sets optomechanical systems apart from 
resonantly coupled light-matter
systems, and it offers time-dependent optical 
control, which is
beneficial in a large range of scenarios, 
including (as we will show) the study of quench physics.

Leaving behind the standard system of one optical mode coupled to one mechanical mode, we arrive at optomechanical arrays (see e.g.
\cite{bhattacharya2008multiple, 
chang2011chang,heinrich2011g, xuereb2012xuereb, 
ludwig2013m, chen2014photon,peano2015topological, zapletal2018dynamically,piergentili2018two,mcdonald2018phase, 
yanay2018reservoir, bemani2019quantum}
). 
These are comprised of a set of coupled vibrational and
optical modes. They can be realized using a variety of 
building blocks, like photonic crystal defect 
cavities or microdisk resonators (in the optical domain), or
microwave-optomechanical circuits. 
Although experimentally still in their infancy 
\cite{zhang2012synchronization,zhang2015synchronization,
fang2017generalized, piergentili2018two,naserbakht2019electro}, 
a variety of promising future directions and 
applications have been identified theoretically, covering
phenomena like bandstructure engineering 
\cite{chang2011chang, schmidt2015optomechanicalDirac}, 
topological transport \cite{peano2015topological,zapletal2018dynamically, mcdonald2018phase}, 
coupling enhancement \cite{xuereb2012strong, xuereb2013collectively,li2016cavity}, 
Anderson localization \cite{roque2017anderson}, 
synchronization \cite{heinrich2011g, holmes2012synchronization}, 
and quantum information 
processing\cite{schmidt2012optomechanical}.

The propagation of photons and phonons in an 
optomechanical array is described by a 
bandstructure of hybrid photon-phonon excitations. 
This bandstructure depends on the geometry 
and the underlying intrinsic coupling of neighboring
optical and mechanical modes. However, on top 
of that, it is also determined by
the external control laser illuminating the array.

In the present work, we demonstrate how 
time-dependent optical control of an
optomechanical array can induce nonequilibrium 
dynamics triggered by non-adiabatic changes in 
the bandstructure.
There are several actively tunable  
degrees of freedom 
in optomechanical arrays that can change the 
bandstructure, e.g. power and phase of the external laser, 
which means that they offer great promise for studying 
non-adiabatic dynamics \cite{schmidt2012optomechanical,brunelli2015out,schmidt2015optomechanical, walter2016classical}. 

In general, nonequilibrium physics produced upon changes of
a Hamiltonian's parameters is encountered 
in many different physical scenarios,
ranging from the evolution of fields in 
the expanding early universe 
to quenches through phase transitions upon 
rapid cooling of a substance 
\cite{polkovnikov2011colloquium, eisert2015quantum, mitra2017quantum}. 
When the parameters of a bandstructure are changed, 
existing equilibrium excitations
will be redistributed. If the quench takes 
the bandstructure through a
topological phase transition, in a finite 
system topological states can be
created or destroyed at the boundaries. 
We will show that this kind of
physics can be explored in optomechanical 
arrays. 
Among our examples of 1D arrays, 
we will present a design for an optomechanical 
Su-Schrieffer-Heeger model \cite{su1979solitons}, 
where 0D edge states exist \cite{atala2013ZakMeasurement, meier2016observation}. 
This model is considered to be the simplest
example of a bandstructure with topological properties
\cite{asboth2016short}.

Quenches through topological phase 
transitions have recently attracted a lot of attention
\cite{sharma2016slow, sun2018uncover, 
schuler2018quench, gong2018topological}.
It is interesting to understand 
how different properties of many-body systems, such as 
integrability \cite{iyer2013exact,
essler2014quench,kormos2017real}
or topological order \cite{tsomokos2009topological,budich2016dynamical,
wilson2016remnant,sharma2016slow} interplay with  
non-equilibrium dynamics of these systems and
how topological properties
such as the Chern number or Berry phase would evolve
through a quench \cite{yang2018dynamical,chen2019linking,  
liou2018quench}. 
For instance, Ciao et al. investigated
some of these questions in the Haldane model 
\cite{caio2015quantum}. Similar investigations 
has been done for the SSH model in cold atoms 
\cite{meier2016observation}.

Although optical lattice experiments 
(like \cite{meier2016observation, 
atala2013ZakMeasurement}) are 
naturally suited for studying quench 
physics and topological phases, we believe our 
work shows it is worthwhile to extend 
such studies to optomechanical systems. 
Not only do they offer different forms of 
access (e.g. via the light emitted from the array), 
but they also involve physics that cannot easily 
be investigated in cold atom systems. 
This includes the effects of a thermal 
environment on the quench dynamics, or 
the possibility to add superconducting 
qubits (in microwave optomechanics 
realizations of optomechanical arrays 
\cite{teufel2008dynamical,regal2008measuring, teufel2011sideband, pirkkalainen2013hybrid}).

The structure of this paper is as follows. 
We start by describing a 1D optomechanical 
array and investigate the quench dynamics 
in this array. This not only helps us 
understand the dynamical properties of this 
particular system, but can also be used for 
other scenarios in which optomechanical 
arrays are driven out of 
equilibrium. Afterwards, we turn to the SSH model. 
After explaining
the basics of the model, we provide a design 
for an optomechanical simulator
that mimics the Hamiltonian of the SSH model and can
also be  tuned dynamically. Finally, we 
describe an example of a simple quench 
experiment that can be carried out using
this simulator and describe the expected outcomes of the
experiment.

\section{Quenches in optomechanical arrays}

\subsection{Model}

\subsubsection{Hamiltonian}

\begin{figure}
\begin{centering}
\includegraphics[width=1\columnwidth]{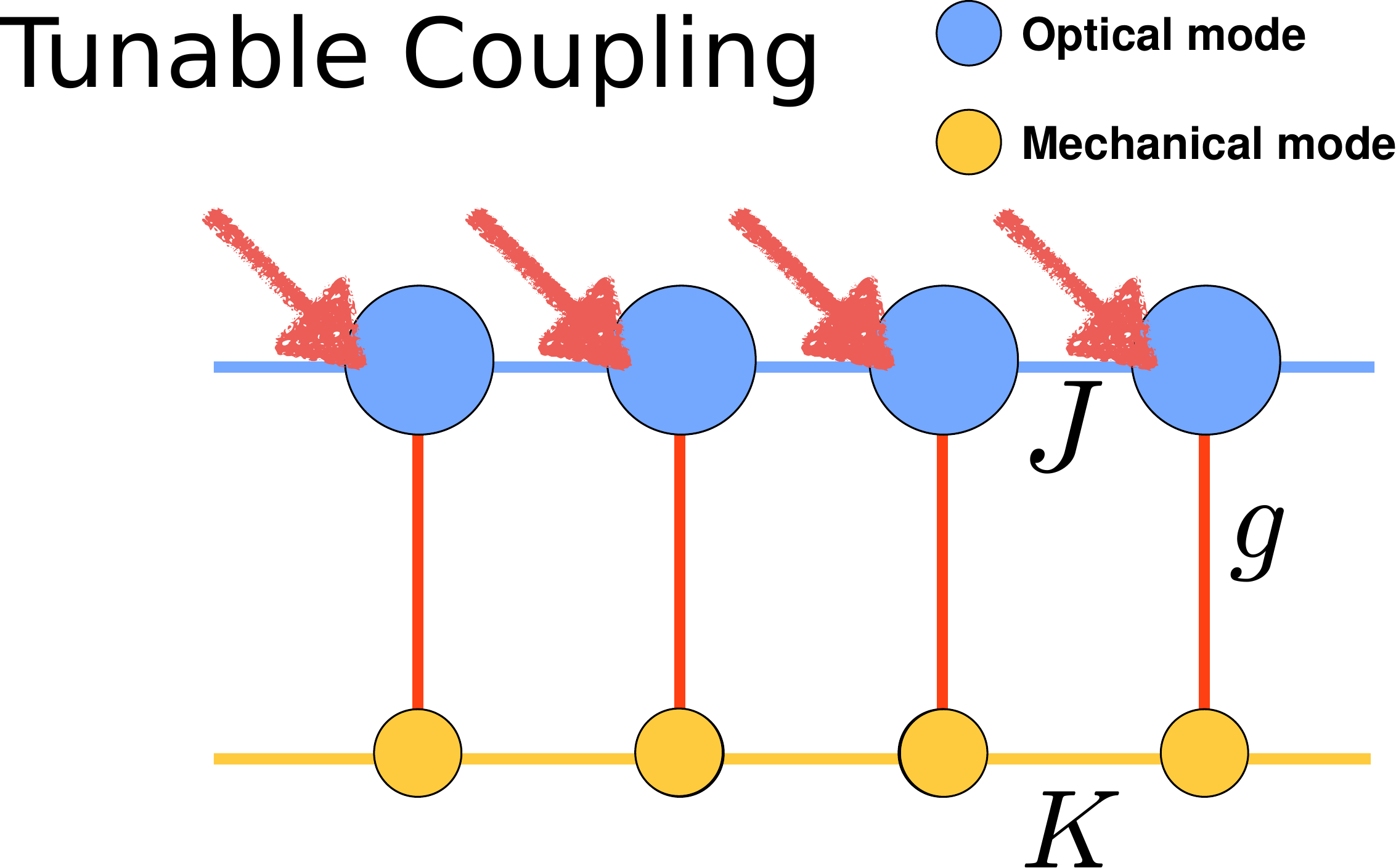}
\par\end{centering}

\caption{\label{fig:Schematic-OMA}Schematic picture of a 1D optomechanical array. 
The blue and yellow circles represent the optical and mechanical modes respectively.
Note that although schematically we separated them spatially, they may occupy 
the same physical space. Blue/yellow links, represent the optical/mechanical
coupling between different sites of the lattice and red links show the optomechanical 
coupling on each site.   
}
\end{figure}

An optomecanical array is an array of optomechanical cells that are
connected through optical and vibrational couplings. 
Figure (\ref{fig:Schematic-OMA})
gives a schematic picture of a simple optomechanical array. 
Blue circles represent the optical cavities with the frequency detuning 
$\Delta$ from the external laser and decay rate $\kappa$. 
Yellow circles represent the mechanical resonators
with the frequency $\Omega$ and dissipation rate $\Gamma$. 
The (laser-enhanced) optomechanical coupling
between the mechanics and optics is given by $g$. 
Furthermore, optical cavities and mechanical resonators on
different sites are coupled to each other and the strength of 
the coupling is given by $\LatK$
and $\LatJ$ for  the mechanical and optical modes between
different sites. The full Hamiltonian of this 
system can be written as
\begin{align}\label{eq:genericOMarray}
H_{OMA} & =\sum_{i}\left(-\hbar\Delta \oprt{a}_{i}^{\dagger}\oprt{a}_{i}+\hbar\Omega \oprt{b}_{i}^{\dagger}\oprt{b}_{i}\right) \nonumber \\
 & -\hbar\sum_{i}g\left(\oprt{a}_{i}^{\dagger}\oprt{b}_{i}+\oprt{a}_{i}\oprt{b}_{i}^{\dagger}\right) \nonumber \\
 & +\hbar J\sum_{i}\left(\oprt{a}_{i}^{\dagger}\oprt{a}_{i+1}+\oprt{a}_{i}\oprt{a}_{i+1}^{\dagger}\right)\nonumber \\
 & +\hbar K\sum_{i}\left(\oprt{b}_{i}^{\dagger}\oprt{b}_{i+1}+\oprt{b}_{i}\oprt{b}_{i+1}^{\dagger}\right).
\end{align}
Here $\oprt{a}_i$ and $\oprt{b}_i$ are the annihilation operators corresponding 
to the optical and mechanical modes on site $i$ respectively. 
Note that this is the linearised Hamiltonian and the Hamiltonian
is quadratic. 
The  array can take any geometrical form. 
The Hamiltonian in Eq. (\ref{eq:genericOMarray})
describes the  lattice given in figure (\ref{fig:Schematic-OMA}).

This system can be experimentally realized, for instance using 
optomechanical crystals \cite{safavi2014two}.

For most of this paper, we consider an ideal system with $\kappa\ll g$, 
which represents the strong coupling regime. 
We also assume that we are working in the red-detuned
regime where the amplification terms in the 
Hamiltonian average out.
Towards the end of this section, we will revisit 
these assumptions and consider the effects of large
cavity dissipation and address how detuning would 
affect our  results.  

For simplicity, we Fourier-transform the Hamiltonian and rewrite it
in terms of pseudo-momentum creation and annihilation 
operators which gives

\begin{align}
H_{OMA}&= 
\hbar  \sum_{k} \left(-\Delta\left(k\right)\oprt{a}_{k}^{\dagger}\oprt{a}_{k}\right)
\cr 
+\hbar  \sum_{k} &\left(
\Omega\left(k\right)\oprt{b}_{k}^{\dagger}\oprt{b}_{k}
+g\left(\oprt{a}_{k}^{\dagger}\oprt{b}_{k}+
\oprt{a}_{k}\oprt{b}_{k}^{\dagger}\right)\right).
\label{eq:HOMA-k}
\end{align}

Here $\Delta\left(k\right)=\Delta-2\LatJ\cos\left(k\right)$ and $\Omega\left(k\right)=\Omega+2\LatK\cos\left(k\right)$.

The Hamiltonian in Eq. (\ref{eq:HOMA-k}) 
is a good approximation for large enough lattices
or periodic BC or the bulk of the lattice, where
there is translational invariance and $k$ is a 
good quantum number. 

The Hamiltonian in Eq. (\ref{eq:HOMA-k}) 
is similar to a single  \OM cell and therefore, 
our results 
can be extended to 
 \OM systems as well.

To study the normal modes of this system, we rewrite the 
Hamiltonian in terms of the Bloch Hamiltonian, $h_k$, i.e.

\begin{equation}\label{eq:HOMA-k-B}
H_{OMA}=\sum_k \hbar\left(\begin{array}{cc}
\oprt{a}_{k}^{\dagger} & \oprt{b}_{k}^{\dagger}\end{array}\right)\blochHk \left(\begin{array}{c}
\oprt{a}_{k}\\
\oprt{b}_{k}
\end{array}\right),
\end{equation}
with 
\begin{equation}\label{eq:blochHk}
\blochHk =\left(\begin{array}{cc}
-\Delta\left(k\right) & g\\
g & \Omega\left(k\right)
\end{array}\right).
\end{equation}
This can be rewritten as 
\begin{equation}
\blochHk = \frac{\Omega\left(k\right) -\Delta\left(k\right)}{2}\mathcal{I}
-\frac{\Omega\left(k\right) +\Delta\left(k\right)}{2}\sigma_z
+g \sigma_x,
\label{eq:blochHk_pauli}
\end{equation}
with $\mathcal{I}$ the identity matrix and $\sigma_x$ and $\sigma_z$ the
Pauli matrices for $X$ and $Z$ respectively.

Diagonalization of the $\blochHk$ gives the normal modes of the Hamiltonian
and the corresponding frequencies. For any value of
$k$, there are two eigenstates which give the normal modes, 
 and we refer to them 
as $\left\{ \Oprtk{A}{k},\Oprtk{B}{k}\right\} $.
These normal modes can be expressed as 
linear superpositions 
of the original modes $\{\oprt{a}_k, \oprt{b}_k\}$, via a unitary transformation 
that diagonalizes the Bloch Hamiltonian, i.e.

\begin{equation}
\left(\begin{array}{c}
\Oprtk{A}{k}\\
\Oprtk{B}{k}
\end{array}\right)=R_k \left(\begin{array}{c}
\oprt{a}_k\\
\oprt{b}_k
\end{array}\right)
\end{equation}

(See the SM for more details.)

For the simulations in this work, we use $\params$  
which are compatible with some of the state-of-the-art 
experiments. See \cite{fang2017generalized} for instance. 

Note that, as an approximation, we only consider dissipation 
for obtaining the initial state, while neglecting 
it during the (fast) quench evolution.
This approximation is valid as long as $\kappa \tau_Q\ll 1$, 
where $\tau_Q$ is the time duration of the quench evolution. 
We will later return to the question of what changes are 
generated by taking into account a finite dissipation rate.
However, considering recent advances in optomechanics and electromechanics 
\cite{groblacher2009observation,teufel2011circuit}, 
this regime should be feasible experimentally.

\begin{figure}
\begin{centering}
\includegraphics[width=\columnwidth]{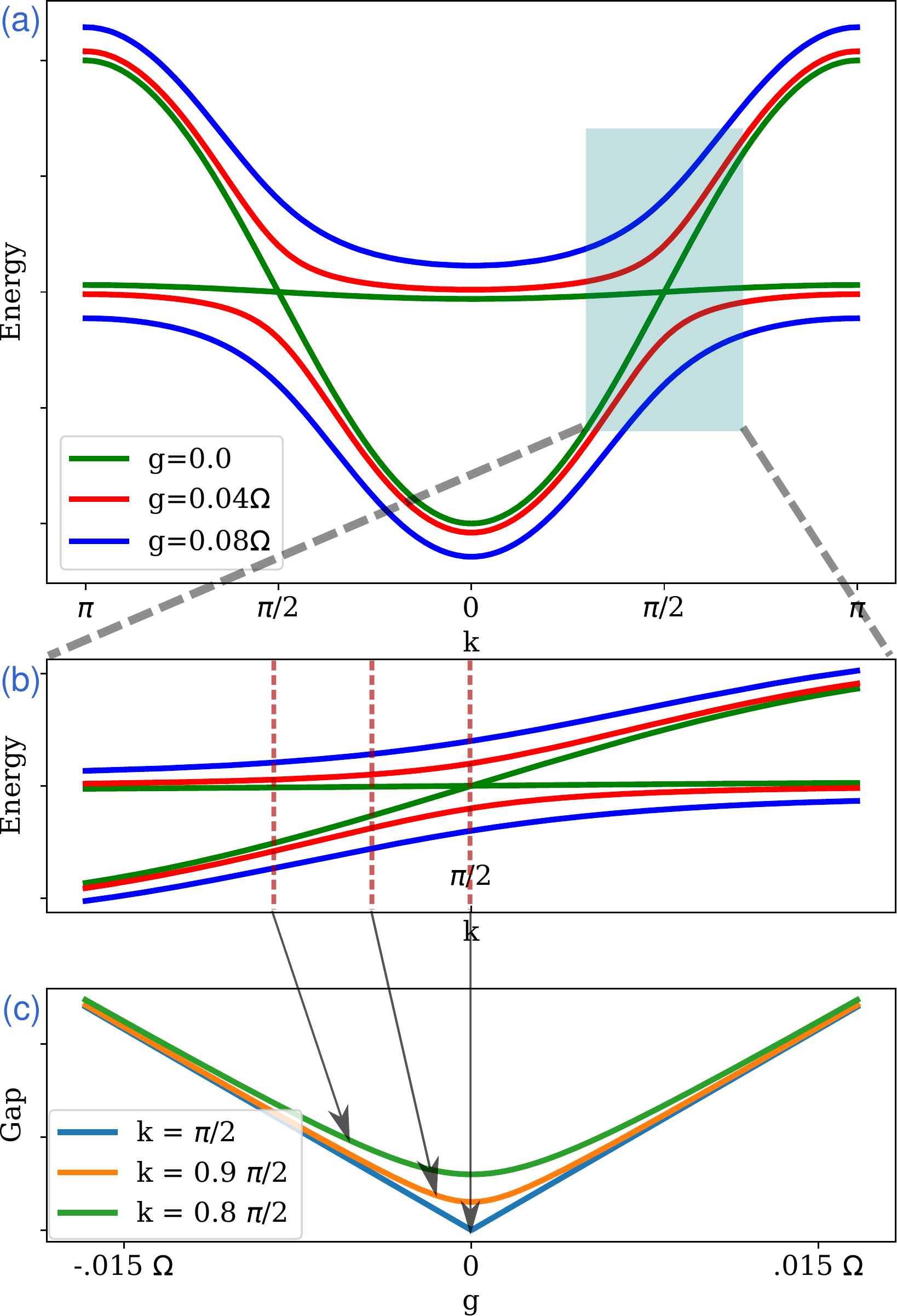}
\par\end{centering}

\caption{\label{fig:Spectrum-SOM-vs-k}
Bandstructure of the \OM array for detuning $\Delta = -\Omega$, 
(a) shows the energy of the modes versus $k$, with different colors 
denoting different values of the laser-tunable optomechanical coupling, $g$.
(b) shows a close-up view of the crossing point at 
$k=\pi/2$. For $g=0$ it is a full crossing whereas in the 
presence of \OM coupling, it turns to an avoided crossing. 
(c) Shows the gap as a function of the coupling $g$ for 
different values of $k$. For $k=\pi/2$ the gap fully closes
at $g=0$. 
}
\end{figure}

Figure (\ref{fig:Spectrum-SOM-vs-k}) 
illustrates the resulting band structures of 
the \OM array for $\Delta = -\Omega$. 
The first term in equation (\ref{eq:blochHk_pauli}) is 
proportional to Identity and only shifts the band structure. 
If we ignore the first term, 
for the regime of $\Delta = -\Omega$, 
the Hamiltonian of the system is 
\begin{equation}
\blochHk = \left(\LatJ-\LatK\right)\cos\left(k\right) \sigma_z
+g \sigma_x.
\label{eq:blochHk_pauli_params}
\end{equation}
Figure (\ref{fig:Spectrum-SOM-vs-k}) shows how the spectrum and also the 
energy gap between the two modes depend on the value of $k$. 
For $k=0$, this gap is the largest and the gap is minimal at 
$k=\frac{\pi}{2}$. 
In the absence of optomechanical
coupling, when $g=0$, the phononic 
band is almost flat and there are two crossings where the gap 
fully closes.
In the presence of optomechanical coupling, 
$\left|g\right|>0$, the two bands do not cross. 
Far from the crossing points and in the middle, mode $\Oprtk{A}{k}$
is mostly phononic and outside, it is mostly photonic. Similarly, mode
$\Oprtk{B}{k}$ is dominated by the photonic mode for $-\pi/2 \leq k\leq \pi/2$
and by phononic modes outside this range. 

From Eq. (\ref{eq:blochHk_pauli_params}), the gap between the two bands 
can be calculated as
\begin{equation}
\Delta_{g} = 2\sqrt{g^2 + \cos\left( k\right)^2 \left( \LatJ-\LatK \right)^2}.
\label{eq:OM_Lat_blk_gap}
\end{equation}

Here we are interested in the dynamical behaviour of the
modes and their population as the Hamiltonian evolves. 
We focus on changing the optomechanical 
coupling. This is done via changing the driving power  of the  laser
and, at each given value of k, takes the Hamiltonian through an
avoided crossing (crossing if $k=\frac{\pi}{2}$) 
and could drive the system out of equilibrium. 

We investigate the excitations from the mode $\Oprtk{A}{k}$ to the
mode $\Oprtk{B}{k}$ as the Hamiltonian evolves through the 
avoided crossing.

\subsubsection{Quench}

We change the coupling $g$ according to
\begin{equation}
g\left(t\right)=g\left(0 \right)(1-\frac{2t}{\tau_{Q}}),
\label{eq:g_change}
\end{equation}

where $\tau_{Q}$ represents the quench time and describes how fast the change
is applied to the Hamiltonian. 
This can be set for instance by the rate at which the external laser changes 
in an experimental 
setting. 
The quench dynamics proceeds from $t=0$ to $t=\tau_Q$, 
switching the sign of the coupling from $+g(0)$ to $-g(0)$.
Large $\tau_{Q}$ describes a slow
change and adiabatic evolution and low 
$\tau_{Q}$ describes a more abrupt evolution. 
We set the time $t$ to start from zero and 
to go to $\tau_{Q}$. 
This makes the Hamiltonian time dependent.  

The range of the time $\tau_Q$ should be set by the band gap in the system in 
Eq. (\ref{eq:OM_Lat_blk_gap}), i.e. for $\tau_Q >\frac{ g}{ \Delta_{(g=0)}^2}$
the evolution would be adiabatic. This limit depends on the value 
of $k$, which means that a specific rate, $1/ \tau_Q$, could be adiabatic for
some values of $k$ and non-adiabatic for the rest of the range. 
For instance, for $k=\frac{\pi}{2}$, the gap fully closes and 
no matter how large the $\tau_Q$ is, the evolution cannot be 
adiabatic.

With the time evolution of the \OM coupling, 
the normal modes would also become time dependent. 
To avoid confusion with the 
time evolution of the modes, we refer to the normal modes 
with respect to their corresponding value of $g$, namely
$\{ \Oprtkg{A}{k}{g}, \Oprtkg{B}{k}{g} \} $ which are calculated from 
the eigenvectors of $\blochHk (g)$ and
\begin{equation}
\left(\begin{array}{c}
\Oprtkg{A}{ k}{g}\\
\Oprtkg{B}{ k}{g}
\end{array}\right)=R_k(g) \left(\begin{array}{c}
\oprt{a}_k\\
\oprt{b}_k
\end{array}\right)
\end{equation}

\subsubsection{Time evolution}

We  use the equation of motions for 
$\{\oprt{a}_k(t),\oprt{b}_k(t)\}$ to find the time propagator of the evolution.
 We break down the evolution to
small enough time-steps. The Hamiltonian should stay constant
over the time-step (compared to $\Vert \blochHk (t)\Vert $). 
Then the time propagator is specified with 
\begin{equation}\label{eq:S}
\frac{dS_k(t)}{dt} = -i h_k(t) S_k(t)
%
\end{equation}
with the initial condition $S_k(0)=\mathcal{I}$ and
 $\blochHk$ is the Hamiltonian in Eq. (\ref{eq:blochHk}) 
and $\delta t$ is the time step. Note that $\delta t \ll 1/ \sqrt{\Vert \left [\blochHk(t),\blochHk(t+\delta t)   \right]\Vert } $. 

The operator $S_k(t)$ gives the 
evolution of the original modes $\{\oprt{a}_k,\oprt{b}_k\}$ as

\begin{equation}
\left(\begin{array}{c}
\oprt{a}_k(t)\\
\oprt{b}_k(t)
\end{array}\right)
=S_k\left(t\right)\left(\begin{array}{c}
\oprt{a}_k\\
\oprt{b}_k
\end{array}\right).
\end{equation}

Now we can calculate the 
evolution of the normal modes too, which is given by

\begin{equation}\label{A(t,t)}
\left(\begin{array}{c}
\Oprtkgt{A}{k}{g(t)}{t}\\
\Oprtkgt{B}{k}{g(t)}{t}
\end{array}\right)
=R_k(g(t))S_k\left(t\right)\left(\begin{array}{c}
\oprt{a}_k\\
\oprt{b}_k
\end{array}\right).
\end{equation}
(See the SM for more details. )

Next we need to specify 
the initial state. 
Each mode could be populated with multiple excitations
and therefore just knowing the evolution of the modes is not enough
to track the excitations.

\subsubsection{Initial state}

One simple choice is to start with a single excitation in one of the
normal modes. It however would be challenging  to 
create a single excitation with a specific momentum experimentally. Probably
the more realistic state to start with  is the thermal state.
This is the stationary
state of the \OM array.  More specifically, we assume that 
before we start changing the Hamiltonian, the system has enough time 
to reach its equilibrium with its environment. 
The normal mode populations of the stationary state are given by

\begin{align}\label{eq:Nin}
n_{th}^{m}\left(\oprt{A}_k\right)= & \frac{\left(1-p_k\right) \Gamma n^M_{th} }{p_k \kappa + (1-p_k)\Gamma}\\
n_{th}^{m}\left(\oprt{B}_k\right)= & \frac{ p_k \Gamma n^M_{th}}{(1-p_k)\kappa + p_k \Gamma}.
\end{align}
where $p_k$ is given by the projection of the normal mode $\oprt{A}_k$ 
on the original mode $\oprt{a}_k$. 

Note that this is assuming that the optical bath is at zero temperature or 
equivalently, $k_B T \ll \hbar \omega_{optical}$. In this regime, we can scale 
the population of the two modes to $1/n_{th}^{M}$ as in Figure (\ref{fig:Pi_Pf_vs_k}).
For more details, see the SM.

\subsubsection{Method}
We  initiate the system in the thermal state
and let it evolve under the time dependent Hamiltonian.
We probe the occupation number of the normal modes 
through the evolution, namely, we look at
\begin{align*}
& \langle\psi(t)\mid\Oprtkg{A}{k}{g\left(t\right)}^{\dagger}\Oprtkg{A}{k}{g\left(t\right)}\mid\psi(t)\rangle\\
& \langle\psi(t)\mid\Oprtkg{B}{k}{g\left(t\right)}^{\dagger}\Oprtkg{B}{k}{g\left(t\right)}\mid\psi(t)\rangle
\end{align*}

See the SM for more details on how we calculate these quantities in 
our simulations. 

\subsection{Results}

\begin{figure}[t]
\begin{centering}
\includegraphics[trim={0cm 0cm 0cm 0cm},clip,width=.83\columnwidth]{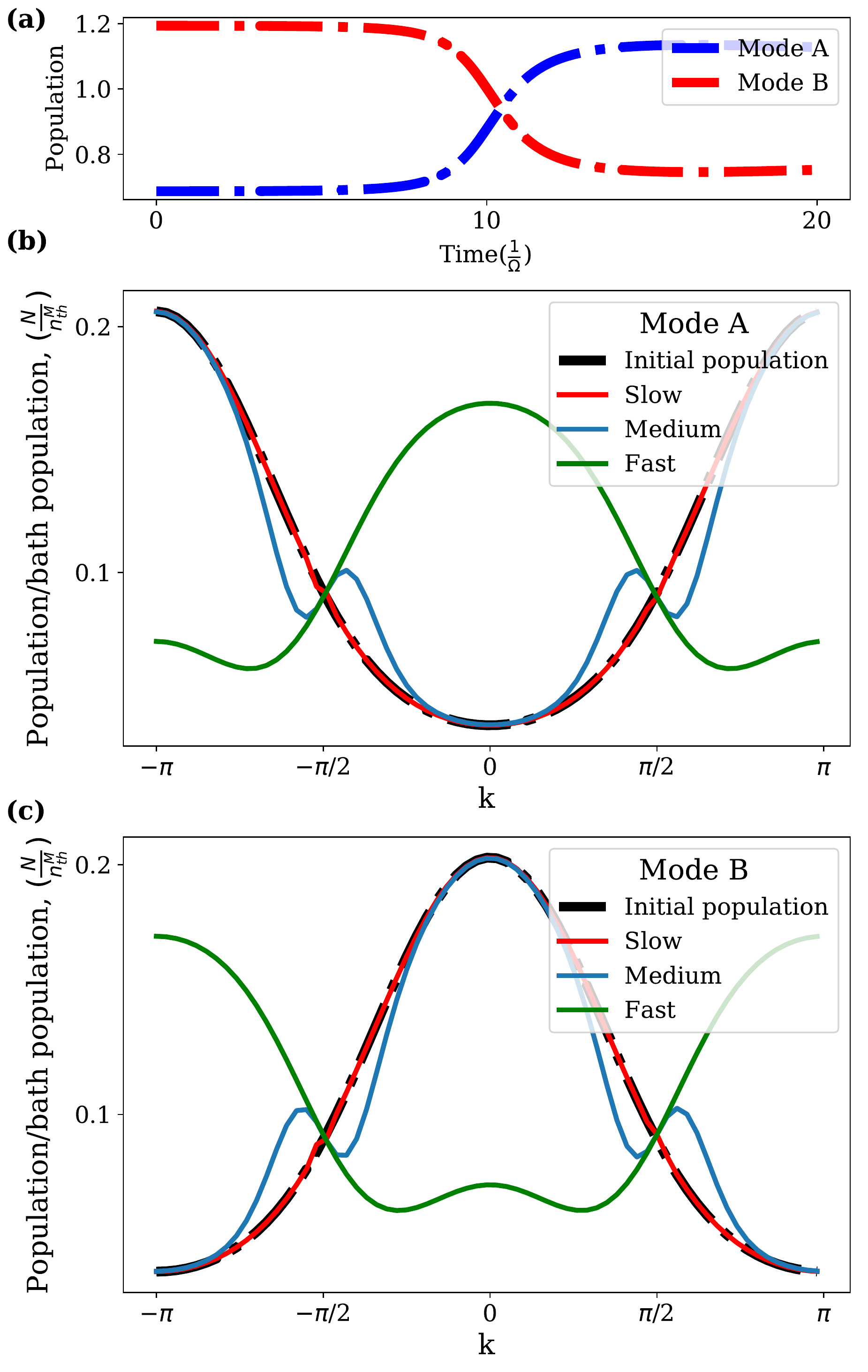}
\par\end{centering}
\caption{\label{fig:Pi_Pf_vs_k}
Quench dynamics for a quench from positive to negative coupling ($g\left(0 \right)$ to $-g\left(0 \right)$).  
(a) shows how the population changes through the quench for 
$k\approx.2\pi$ and some medium speed quench. 
(b) and (c) show the final population scaled to the thermal population of the bath, 
$N_f/n_{th}^{M}$, 
 of mode $A$ (upper branch) and likewise mode B (lower branch)  
after a slow, medium-speed and fast quench. 
For the slow, medium and fast quench, 
$\frac{\tau_Q}{g(0)/\Delta_{g=0}^2} \approx.0001, .01, 1$ 
respectively.
The initial population $N_i$
is also included for comparison. The initial population $N_i$
is calculated based on the thermal equilibrium state with 
a bath and the final population is derived evolving 
the initial state while changing the Hamiltonian. 
Depending on the rate at which the Hamiltonian changes, 
the overall evolution can be non-adiabatic or adiabatic. 
For the slow quench, the final population, $N_f$, is close to the initial
one, however for the fast and mid-speed quenches, 
the evolution generates some excitations.
The band structure and the gap between 
 the two bands depends on $k$, so the final 
populations $N_f$ would also depend on  $k$. 
For instance the gap closes for $k=\pi/2$ and both modes
would have the same energy, so no matter 
how slow we quench the Hamiltonian,
in the vicinity of this point, 
the evolution would always be non-adiabatic 
and the value of $N_f$ would increase. }
\end{figure}

We start by comparing a fast and a slow quench. Figure (\ref{fig:Pi_Pf_vs_k})
shows the simulation results for the final excitations for a slow, mid-speed 
and a fast quench.
The top plot shows the results for excitations in mode A and the bottom one 
shows the excitations in mode B. 
For comparison, we included the initial population given by Eq. (\ref{eq:Nin}). 
It is critical to take 
these initial excitations into account when we study the excitations generated by
the quench. 
Figure (\ref{fig:Pi_Pf_vs_k}) also shows
that for a slow quench, the number of excitations stays almost unchanged,
whereas for the fast quench, new excitations are generated through
the quench process. 

Provided we assume $\Delta=-\Omega$ 
(as we will do for these simulation), the gap between
the two bands, $\Delta_g$, 
vanishes for $k=\pi/2$  
(See Eq. (\ref{eq:OM_Lat_blk_gap}) and figure (\ref{fig:Spectrum-SOM-vs-k})) and as a
result, the dynamics is always non-adiabatic 
at this point. This explains why there
are excitations generated in the vicinity of $k=\pm\pi/2$, even for 
the slow quench.

Here we focus on the net excitation,
$N_{Q}$ which is
\begin{equation}
N_{Q}=N_{f}-N_{i},
\end{equation}
where $N_{f}$ and $N_{i}$ represent the final and initial population of the bands.

\begin{figure}
\begin{centering}
\includegraphics[width=\columnwidth]{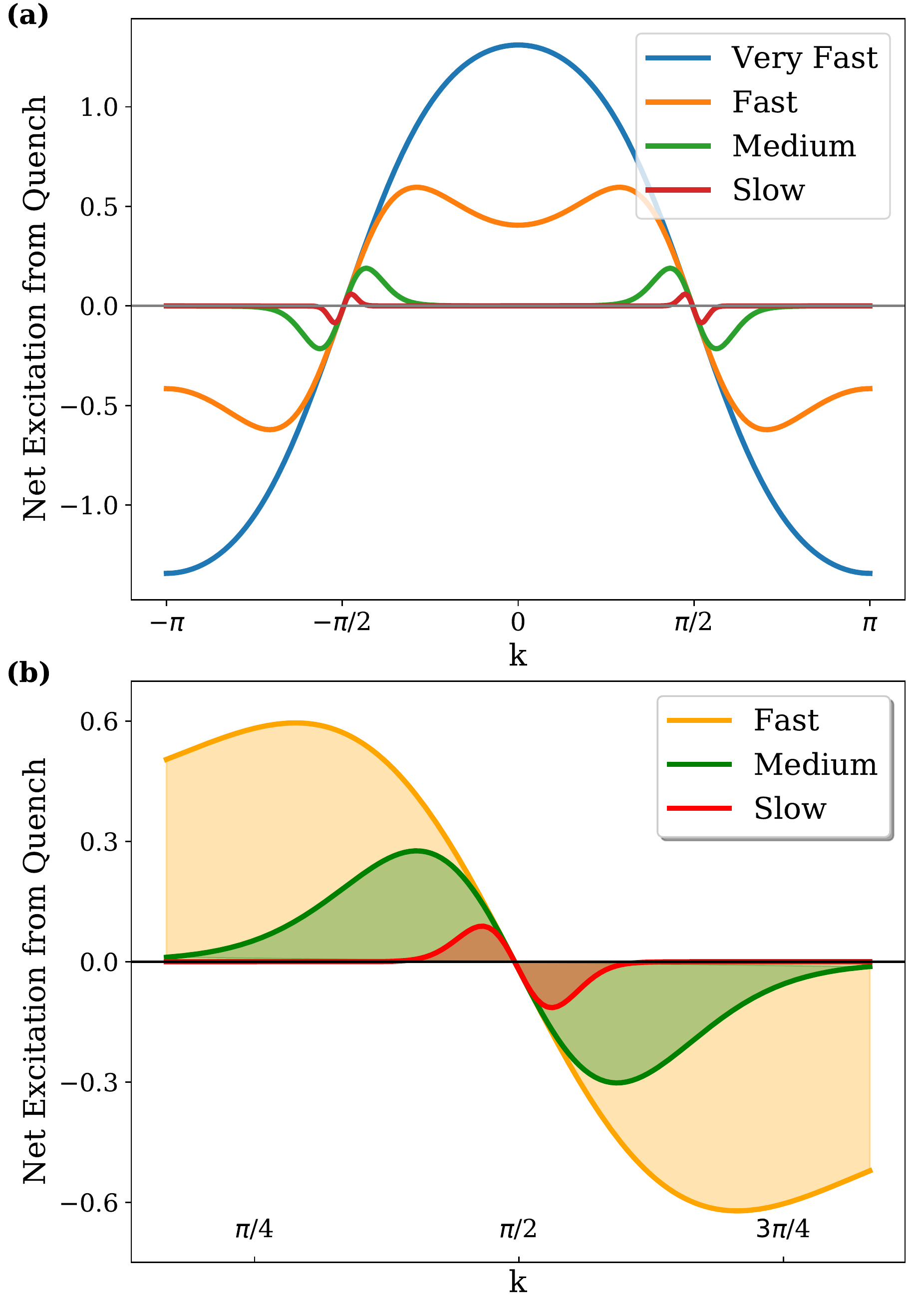}
\par\end{centering}

\caption{\label{fig:Nf-Ni_vs_k}
Net excitation in mode A for different quench times, $\tau_{Q}$. 
Different plots 
represent different quench times. 
From the top
to the bottom, $\tau_Q$ increases. 
For the slow, medium, fast and very fast quench, 
$\frac{\tau_Q}{g(0)/\Delta^2_{g=0}} \approx .03, .01, .001, .0001$ 
respectively.
This indicates that the slower the quench, 
the less excitation would be generated and 
the smaller the non-adiabatic regime would be. 
(b) shows a close-up view of the plot in (a) around 
$k=\pi/2$, where the gap closes. 
This indicates that even for slower quenches, 
there are some excitations around the level-crossing point. 
For the simulations here we used $\params$. 
See the text for more details. }
\end{figure}

Figure (\ref{fig:Nf-Ni_vs_k}) shows the net excitations in mode A for different
quench times, $\tau_{Q}$. This figure indicates that there is a regime
for which the dynamics is non-adiabatic. We introduce $\NAdk$
to indicate the range of the non-adiabatic regime. 
We define $\NAdk$ as the maximum distance from 
$k=\frac{\pi}{2}$ where the excitation generated by the quench, $N_{Q}$, 
is above some threshold $\epsilon$. 
Note that there are two non-adiabatic regions, one around $k=\frac{\pi}{2}$ and
one for $k=\frac{-\pi}{2}$
We only consider the region around $k=\frac{\pi}{2}$ 
for simplicity and restrict our discussion to positive values of $k$. 
Mathematically, that is
$\NAdk = \min \{p \mid \forall k \mid k - \pi/2 \mid >p,N_{Q}\left(k\right)<\epsilon\}$
where $\epsilon$ is some threshold. The parameter $\NAdk$ is mostly affected by
the quench time $\tau_{Q}$. 

\begin{figure}
\begin{centering}
\includegraphics[width=\columnwidth]{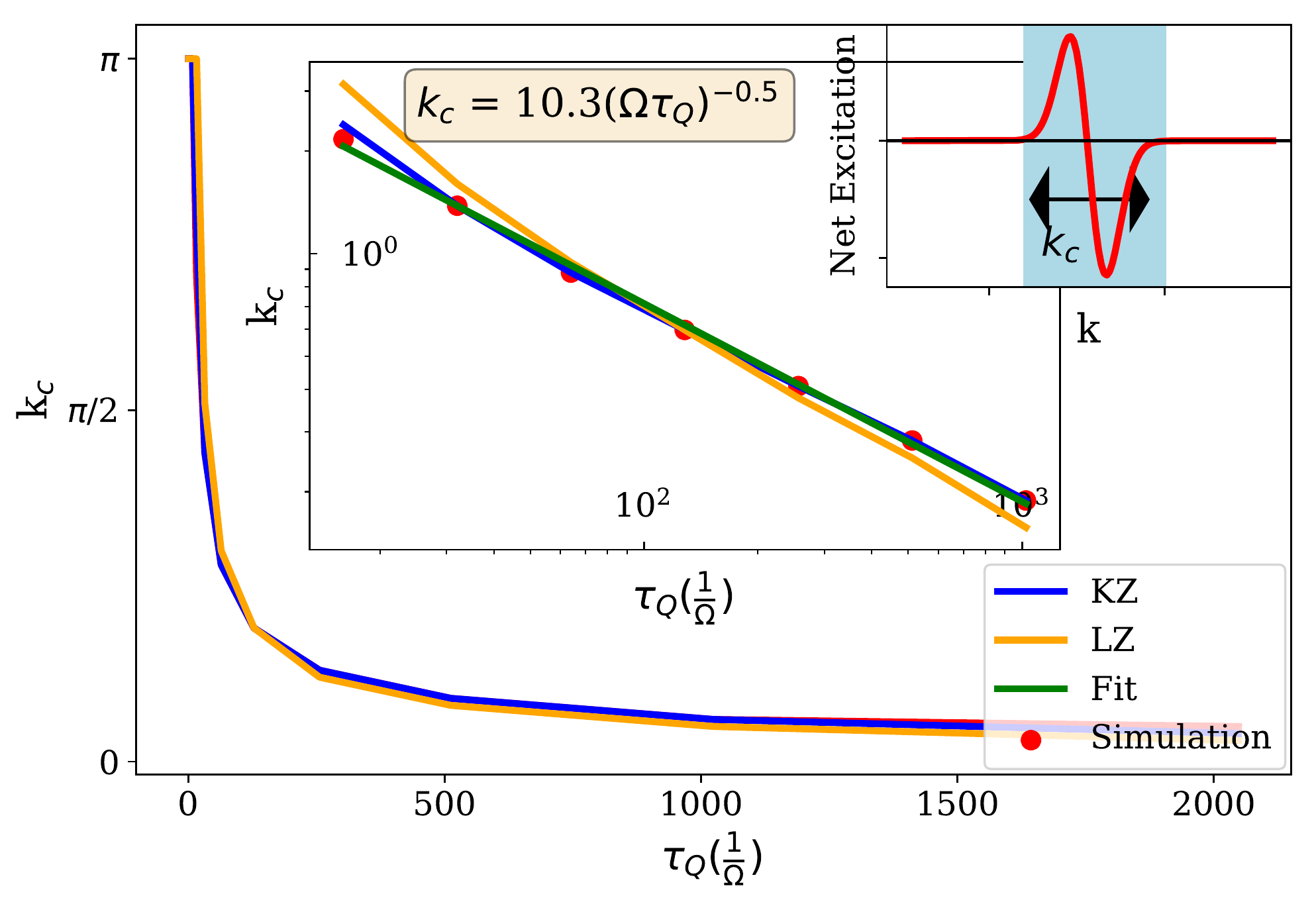}
\par\end{centering}

\caption{\label{fig:k*-vs-Tq}The extent of the non-adiabatic regime, $\NAdk$,  as a function
of quench time, $\tau_{Q}$. This plot illustrates a power-law dependence on the
quench time.  
We included the two analytical estimates along the best linear fit
for the simulation results which 
are in good agreement. 
The first analytic estimate comes from the Landau-Zener
formula. The second one is adopted based on the Kibble-Zurek
mechanism. See section (\ref{Sec:AnaAsses}).
The fit and its corresponding equation are included in 
the log-log plot in the inset in the middle. 
Note that fit is expected to give 
$ \NAdk\approx\frac{\sqrt{2g(0)}}{(J-K)\sqrt{\pi\tau_{Q}}}$, 
that is mostly affected by the values of $g(0)$ and $J$ and for the parameters here gives $\NAdk\approx\frac{10}{\sqrt{\tau_{Q}}}$
The top-right inset shows the non-adiabatic regime in the 
net excitation plot from figure(\ref{fig:Nf-Ni_vs_k}-b). 
}
\end{figure}

Figure (\ref{fig:k*-vs-Tq}) shows  $\NAdk$ as a function of the
quench time, $\tau_{Q}$. This plot shows the power-law dependence
of the size of the non-adiabatic regime, $\NAdk$, on the quench time.  

For these simulations, we take the following values
for the quench time
\[\tau_Q \in \frac{50}{\Omega}\times\{2^{-1}, 2^0, 2^1, \cdots, 2^{10}\}. \] 
These values are set such that the smallest value would give 
a non-adiabatic evolution for all values of $k$ and
the largest value 
would give an  
adiabatic evolution for essentially all the values of $k$ that we consider 
in our simulations.

Next we will assess the dynamics analytically and 
show that these results are compatible with analytical expectations.

\subsection{Analytical assessment \label{Sec:AnaAsses}}

The simulation results here can be approximated with the Landau-Zener (LZ) formula
for excitations in a time-dependent two-level system. For a two-level 
system with $\left( h_k\right)_{12} = (J-K)cos(k)$ as the off-diagonal elements of 
the Hamiltonian, the Landau-Zener formula 
\cite{zener1932non,landau1958quantum}
gives
the probability of excitation as  
\begin{equation}
P_{LZ}= 
e^{\frac{\pi \left(\LatJ-\LatK
\right)^2}{2 g\left(0 \right)}\cos\left(k\right)^{2}\tau_{Q}}.
\end{equation}
 Note that we used the Hamiltonian in Eq. (\ref{eq:blochHk_pauli_params})
 to calculate the probability. 
This shows that the border between the adiabatic and non-adiabatic regime is
approximately given by $\tau_{Q}\approx 2g\left(0 \right)/\pi\left(\LatK-\LatJ\right)^{2}\cos\left(k\right)^{2}$, 
i.e. 
if the quench happens on a faster time-scale, 
then 
the evolution would be non-adiabatic and generates excitations and 
similarly, if it is slow, then the evolution would be adiabatic and 
gives no extra excitations.

If we expand this in terms of small $\delta k$ 
from $\pi/2$, 
we have $\cos\left(\pi/2+\delta k\right) = \sin\left(\delta k\right) \approx \delta k$
and we get $ \NAdk\approx\frac{\sqrt{2g(0)}}{(J-K)\sqrt{\pi\tau_{Q}}}$, which indicates 
that the size of the non-adiabatic region in k space, $ \NAdk$, 
has a power-law dependence on the quench time, $\tau_{Q}$. 
The Landau-Zener fit is included in Figure
(\ref{fig:k*-vs-Tq}) for comparison and 
 confirms the simulation results. 

A more intuitive approach is to break down the evolution into two phases, 
the adiabatic and freeze-out zone. This is similar to the  
Kibble-Zurek mechanism (KZ)
\cite{kibble1976topology,zurek1985cosmological,
zurek1996cosmological, nalbach2015quantum}. 

We assume that the dynamics  in the adiabatic
zone is fully adiabatic. Similarly, we assume 
that the state does not change in the freeze-out 
zone. 
Clearly, this is an approximation and 
the transition from adiabatic to non-adiabatic dynamics is usually gradual and the
state does not fully freeze. However, this gives a good fit
 to our numerical simulations. 

Assume that the evolution starts in $t_{i}=-\infty$ 
with the coupling $g(t_i=-\infty)= - g_m$ and goes to
$t_{f}=\infty$ with coupling $g(t_f=\infty)= g_m$, and  that we start with the ground state.  
We use the $\{ \ket{\gs(t)} ,\ket{\ex(t)} \}$ to represent the ground and excited states of 
the Hamiltonian at time $t$. This is not to be confused with 
the optomechanical coupling $g\left( t\right)$.
Note that, for simplicity, we are taking time to symmetrically evolve
from $-\infty$ to $\infty$ which is slightly different 
from our convention in Eq. (\ref{eq:g_change}), but it does not change the result and it can be easily transformed to the convention in Eq. (\ref{eq:g_change}).

More importantly, we assume
that at some time, $-\hat{t}$, the evolution transits from adiabatic
to the freeze-out zone and then becomes adiabatic again at $\hat{t}$.
Under these assumptions, the state evolves as follows
\begin{align*}
\ket{\psi_{i}} & =\mid \gs\left(-\infty\right)\rangle
\rightarrow \ket{\psi\left(-\hat{t}\right)} \approx\mid \gs\left(-\hat{t}\right)\rangle
\rightarrow \\
\ket{\psi\left(\hat{t}\right)} &\approx\mid \gs\left(-\hat{t}\right)\rangle=\alpha\mid \gs\left(\hat{t}\right)\rangle+\beta\mid\ex\left(\hat{t}\right)\rangle\\
\rightarrow & \ket{ \psi_{f}\left(\hat{t}\right)}\approx\alpha\mid \gs\left(\infty\right)\rangle+\beta\mid\ex\left(\infty\right)\rangle.
\end{align*}
First, we start with the state at $t=-\infty$. Up to $t=-\hat{t}$ the evolution
is adiabatic which keeps the state in the ground state. 
From this point, up to $t = \hat{t}$ the state freezes and stays unchanged. 
So at time $t = \hat{t}$, we still have the $\ket{\gs(-\hat{t})}$, which no longer 
represents the ground state, but some superposition of both the ground
and excited states. Beyond this, the evolution is adiabatic again which 
preserves the superposition.

Therefore, the amount of excitations are given by $\left|\beta\right|^{2}$.
In order to calculate $\beta$, we only need to know the projection 
of eigenstates at $-\hat{t}$ to the eigenstates at $\hat{t}$. 

The eigenvectors  of the optomechanical array 
can be calculated from Eq. (\ref{eq:blochHk_pauli_params}) and 
would give 

\begin{equation}\label{eq:Beta}
\mid\beta\mid^{2}=\frac{\left(g_{m}\frac{\hat{t}}{\tau_{q}}\right)^{2}}{\left(g_{m}\frac{\hat{t}}{\tau_{q}}\right)^{2}+\delta^{2}}.
\end{equation}

Next we need to find $\hat{t}$. If we follow   the same idea
as in the Kibble-Zurek mechanism, this is the time at which it takes the same
amount of time for the system to relax as it has to get 
to the crossing point, i.e. 
$\hat{t}=\tau_{\text{relx}}=\frac{1}{\sqrt{\delta^{2}+\left(g_{m}\frac{\hat{t}}{\tau_{q}}\right)^{2}}}$, 
with $\tau_{\text{relx}}$ the relaxation time.
Note that this is not an actual relaxation time, but the 
time scale given by the $\frac{1}{\text{Gap}}$. 

If we plug this into Eq. (\ref{eq:Beta}),  we get
\begin{equation}\label{eq:Beta-f}
\beta = 1-\frac{2\delta^{2}\text{\ensuremath{\tau_{q}}}^{2}}{\delta^{2}\text{\ensuremath{\tau_{q}}}^{2}+\sqrt{\delta^{4}\text{\ensuremath{\tau_{q}}}^{4}+32g^{2}\tau_{q}^{2}}}.
\end{equation}

The KZ analytical fit is also included in figure
(\ref{fig:k*-vs-Tq}) which shows that both analytical assessments are in good agreement 
with the simulation results. 

This concludes the results in this section. We studied the excitations
generated through the quench and showed that they are compatible with
KZ and LZ predictions. 


\subsection{Experimental Imperfections}

Now we investigate the experimental 
challenges of implementing and testing our results. 

As we stated before, we assume that we are working 
in the strong coupling regime, i.e. $g \gg \kappa $. 
This has already been achieved experimentally in 
\cite{teufel2011circuit,groblacher2009observation}.

\begin{figure}
\begin{centering}
\includegraphics[width=.97\columnwidth]{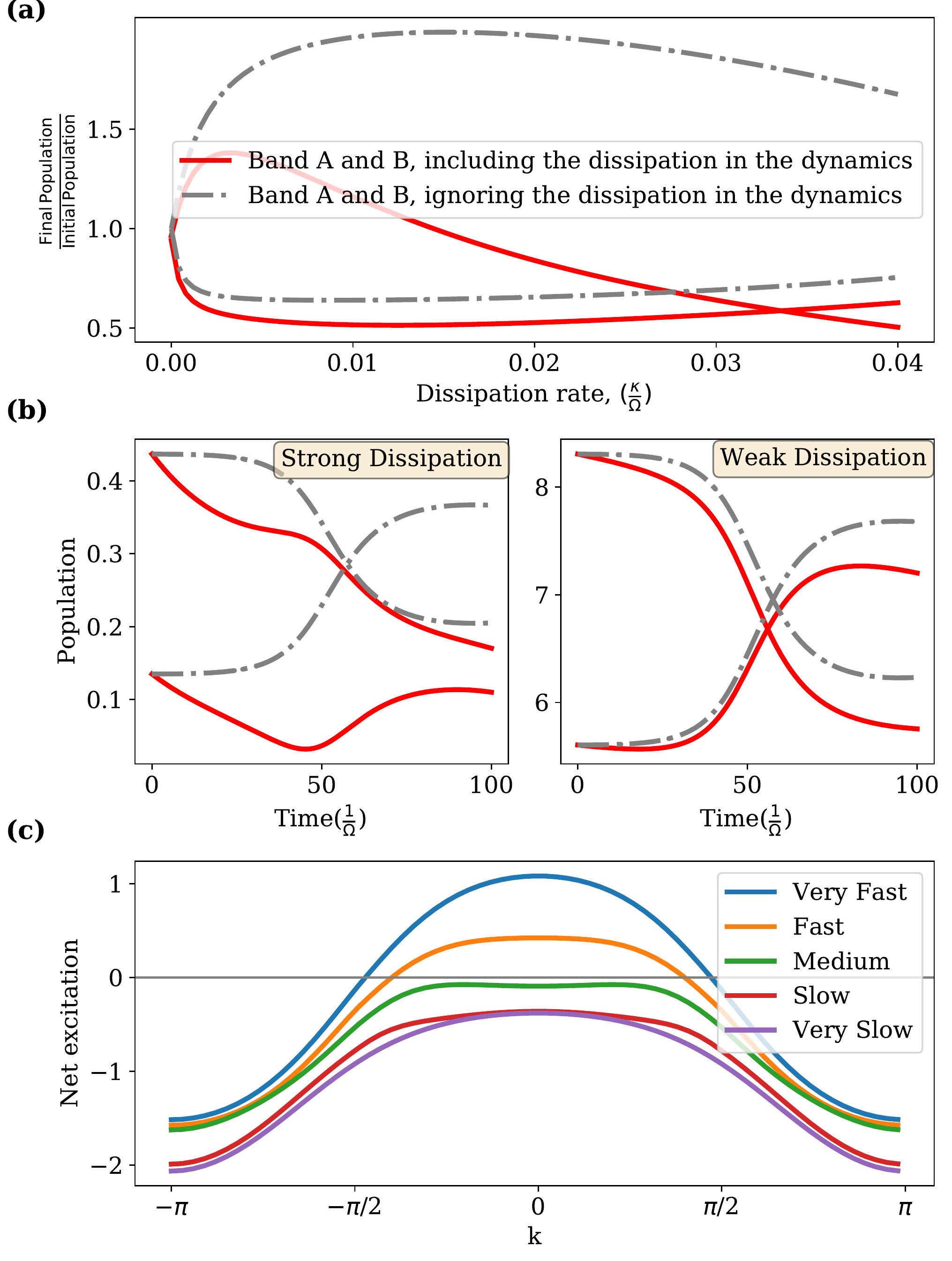}
\par\end{centering}

\caption{\label{fig:CrossingwDissipation}
The influence of dissipation on the evolution and 
final population of the modes.
(a) shows how the final population of the two modes
would be affected in the presence of dissipation. To give
a reference for comparison, the results in the absence of 
dissipation are also included.
Dissipation reduces the final population. 
Note that, since for different amounts of dissipation, 
the initial population changes, the results are normalized to the 
initial values for each point. 
Plots in (b) illustrate 
how the dissipation affects the dynamics of the populations. 
Line colors and styles are the same as the ones in (a). 
In each plot, the 
population of the two modes A and B, both in the presence and absence of 
dissipation, are shown versus time, 
as the system evolves through the avoided crossing. 
For the plot on the right, the quench is still
fast enough for the two populations to cross, 
however, for stronger dissipation, the population of 
 both bands could decay before the can cross. 
\label{fig:DNwDissipation} (c)  shows the 
net excitation generated through the evolution in the 
presence of dissipation. The 
general trend is similar to the one in figure 
(\ref{fig:Nf-Ni_vs_k}),
however, due to the dissipation, the net excitation has dropped. 
Different plots corresponds to different quench times.
For the very slow, slow, medium, fast and very fast quench, 
$\frac{\tau_Q}{g(0)/\Delta^2_{g=0}} \approx .03 .015, .007 ,.0005, .0001$ 
respectively.}
\end{figure}

We also ignored the dissipation for the most part, but
we can also extend our simulation to the situation where 
the dissipation is not ignored. Figure (\ref{fig:CrossingwDissipation}) 
shows how the typical behaviour of this system changes as 
we add dissipation. Without dissipation, the gray
plots show how evolving the Hamiltonian through the 
avoided crossing would swap the populations of the two
modes. However, when dissipation is included, both populations
start to decline to a point that if the quench is not fast enough, 
they would not cross.
Figure (\ref{fig:DNwDissipation}) shows how dissipation would affect the net excitation generated 
through the dynamics. Although the general trend is preserved,
the net excitation is decreased compared to the one in figure (\ref{fig:Nf-Ni_vs_k}).
Note that here we assume that the photonic bath is at zero temperature
which is consistent, considering that typically
 $\hbar \omega_{\text{Optical}} \gg k_b T$, with $\omega_{\text{Optical}}$ the optical frequency. 
 We also assume that $\Gamma \ll \kappa,g$
 for this plot which can be fulfilled in most experiments. 

In all the illustrations so far, we assumed $\Delta = -\Omega$ 
(red detuned regime) and all the mode dynamics to be described 
by the beam-splitter Hamiltonian (which relies on $J,K<<\Omega$). I
n principle, one can consider arbitrary detunings, 
including those where excitations may be generated by the
amplification terms
in the Hamiltonian.  

One of the main challenges in analysing a regime including 
photon-phonon pair generation would be that it is not possible 
to distinguish  the excitations that are generated directly 
by the parametric terms 
from the ones generated by the quench.  
This explains why we focus on the   
regime where number-non-preserving 
terms in the Hamiltonian are suppressed and all the excitations
can be associated to the quench.

\begin{figure}
\begin{centering}
\includegraphics[width=\columnwidth]{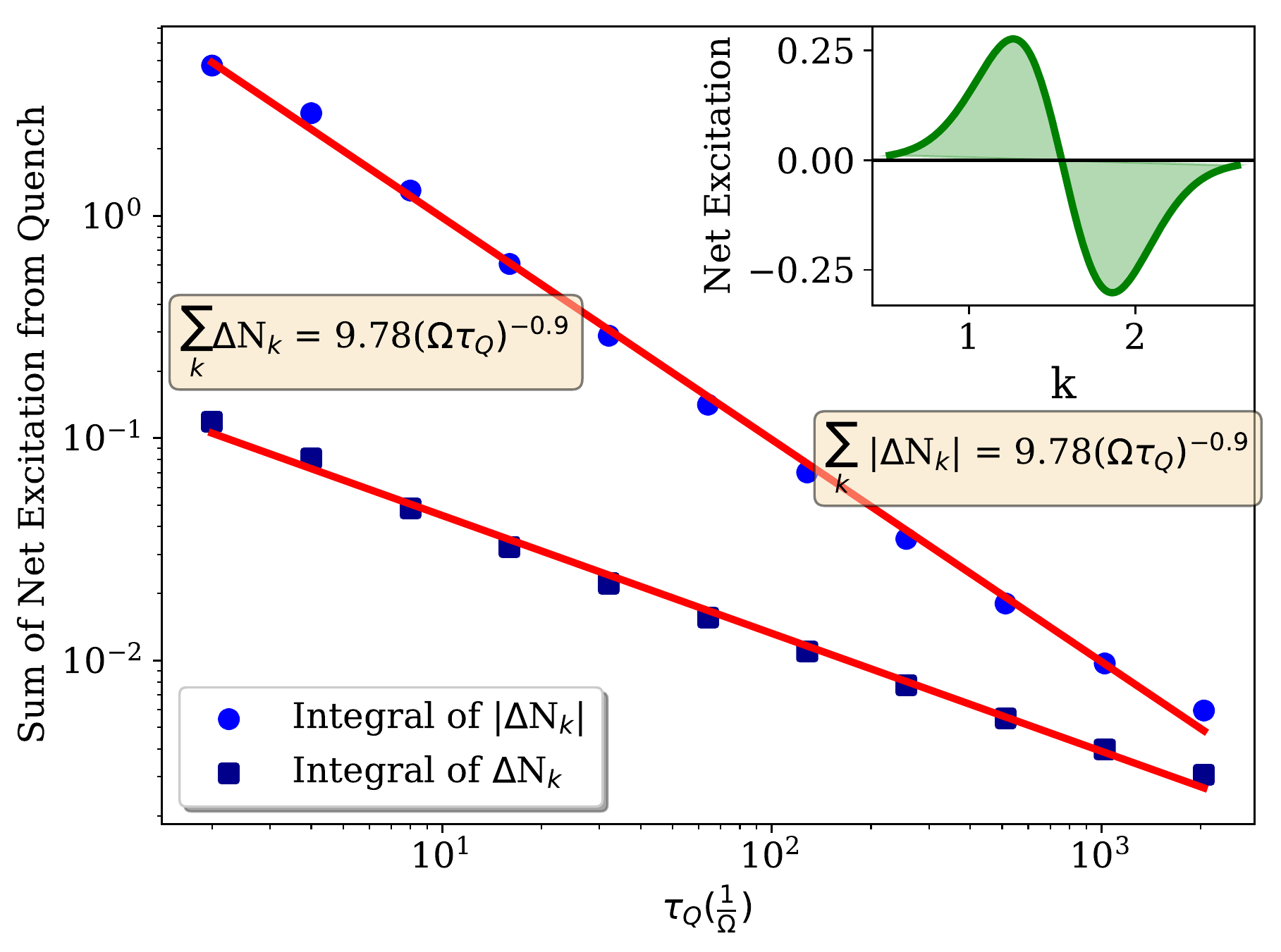}
\par\end{centering}

\caption{\label{fig:Ntotal}
This plot shows the sum of all excitations 
with different momenta as a function
of the quench time. 
The circles show the sum of the absolute values 
of the net excitations and the squares
show the sum, i.e. the integral under the 
plots in figure (\ref{fig:Nf-Ni_vs_k}) for different 
quench times. 
The red curves give the best linear fit to the data points. 
The equations of the fits are included next to the plots. 
The inset shows the net population after the quench
around $k=\frac{\pi}{2}$. See figure 
(\ref{fig:Nf-Ni_vs_k}-b) for more details}
\end{figure}

Another challenge is that for the results   
in figure (\ref{fig:Nf-Ni_vs_k}), excitations with
different pseudo-momentum should be resolved. While this 
is in principle possible \cite{schmidt2013optomechanical}, 
a simpler solution is to look at the sum of the net excitations,
i.e.  $\int N_{Q}\left(k\right)dk$. This is the area under the plot of
$N_{Q}\left(k\right)$.
Figure (\ref{fig:Ntotal})
shows this quantity for different quench times. 
Although the net exitations still follow a power-law, 
the values are too small and probably challenging to detect experimentally. 
Alternatively, we can investigate the absolute value of the net excitation, 
which still gives a power-law, but this would require 
$k$-resolved measurements of the excitations too.

The last assumption that needs clarification is 
the periodic boundary conditions on the lattice,  
which makes it possible to work in Fourier space. It is possible
to do this calculations for a finite-size system and work out the
excitations for different sites on the lattice, but it is computationally
more challenging.

\section{Application: Quenches in  the Optomechanical \SSH $\,$ model} 

So far, the main focus has been to understand how changes
in the Hamiltonian would affect the dynamics of \OM arrays. 
In this section we will give an example to illustrate how \OM arrays can be 
designed to mimic the evolution of the SSH model.
This model exhibits a topological phase transition, which 
makes it a nice candidate for exploiting the dynamical properties of 
the \OM array for simulation purposes.

\begin{figure}
\begin{centering}
\includegraphics[width=\columnwidth]{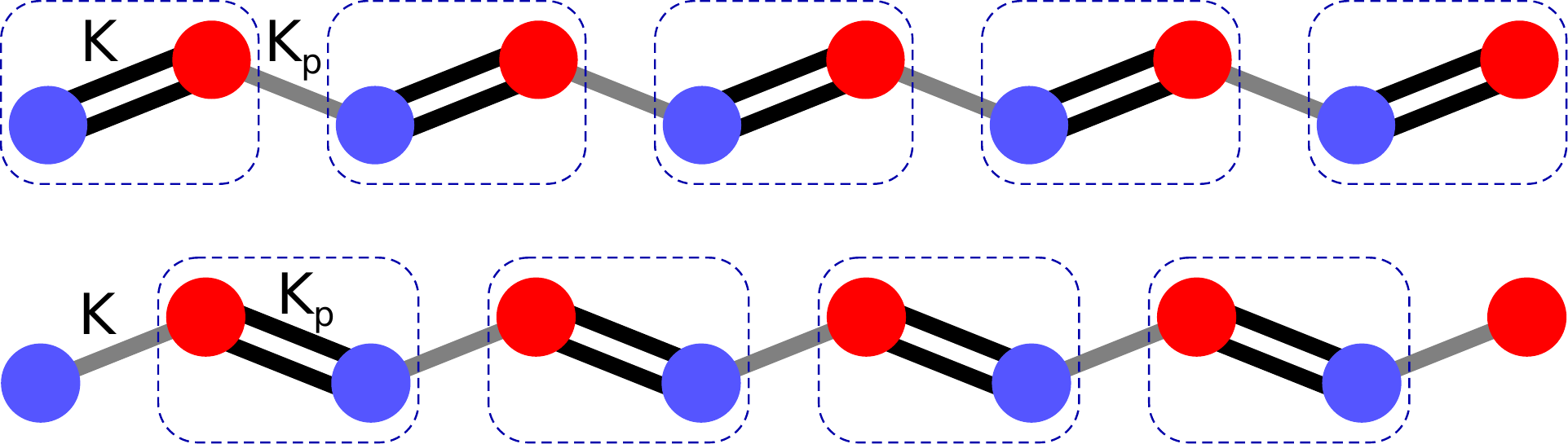}
\par\end{centering}

\caption{\label{fig:Schematic-SSH}Schematic picture of the SSH model. The two figures 
illustrate the two phases of the model, the top one is when $K$ is the 
dominant coupling and bottom is revered. For the latter, 
in contrast to the top one, not all the sites can pair up and two sites 
are left at the two ends of the lattice which produce the zero-energy edge states. }
\end{figure}

The SSH  model describes
a one-D topological insulator \cite{asboth2016short} where fermions can hop from one site to the other,
however, hopping rates are staggered and the hopping rate to the left and right
are different for each site. See figure (\ref{fig:Schematic-SSH})
for a schematic picture of the SSH model. 
The SSH model has two phases that are separated by a topological phase
transition. For the finite size model, one phase exhibits 
zero-energy edge states. 

Here we first present a brief introduction to the SSH model and then
propose an \OM array design that emulates the SSH model and show
how the effective dynamics is compatible with SSH.

\subsection{SSH model}

For the purposes of this work, it suffices to understand the Hamiltonian
and the phase diagram of the SSH model. This model comprises a chain that 
can be separated into two distinct sublattices. We refer to these sublattices
as sublattice A and B. 
Fermions on each sublattice have similar right and 
left hopping rates. This means 

\begin{equation}
H_{SSH}=K \sum_{i=1}^{N}  \left(\oprt{c}_{i}^{\dagger}\oprt{d}_{i}+
\oprt{c}_{i}\oprt{d}_{i}^{\dagger}\right)+
K_{p}\sum_{i=1}^{N-1}\left(\oprt{c}_{i+1}^{\dagger}\oprt{d}_{i}
+\oprt{c}_{i+1}\oprt{d}_{i}^{\dagger}\right),\label{eq:SSHH}
\end{equation}
where $\oprt{c}_{i}$ and $\oprt{d}_{i}$ are the annihilation operators on odd and 
even sites,  
corresponding to the two sublattices. 
Note that to avoid confusion with the creation operators for 
the photonic and phononic modes in the first part of the paper, we 
use $\oprt{c}_{i}$ and $\oprt{d}_{i}$ here.

Here we assume a finite-size lattice with
$2N$ sites. 

\begin{figure}
\begin{centering}
\includegraphics[width=0.9\columnwidth]{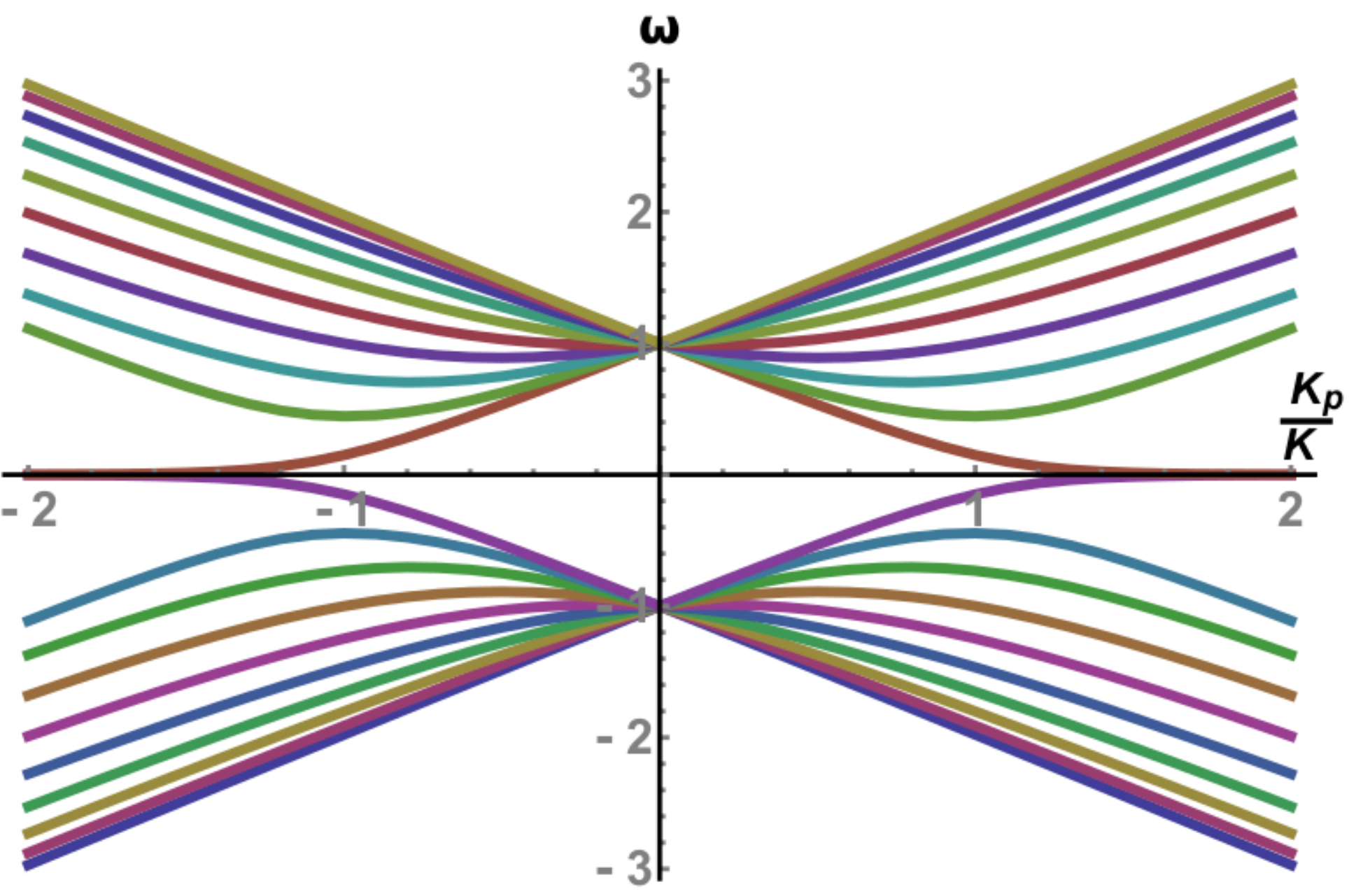} 
\par\end{centering}

\caption{\label{fig:SpectrumSSH} The spectrum of the finite-size 
SSH model with 20 sites and 10 unit cells as a function of the 
ratio of the hopping rates $\{\SSHJ, \SSHJP\}$. For $|\SSHJP|>|\SSHJ|$
two energy levels start to converge and 
form the two zero-energy edge states.  }
\end{figure}

The spectrum of the SSH model with 20 sites (10 unit cells) 
is shown in figure (\ref{fig:SpectrumSSH}). Each line represents one
energy level and the plot shows how energy levels change with the ratio of the
hopping rates. 
This model has two phases which are distinguished by the order parameter
$\lambda=\SSHJP/\SSHJ$. For $\lambda<1$, all the sites pair up and
form dimers. In the opposite regime, i.e. $\lambda>1$, all the particles
in the middle pair up, however, there are two sites left out at the two
ends. 
These two make the two zero-energy edge states of the SSH model.
These are the two energy levels at zero energy in figure 
(\ref{fig:SpectrumSSH}) which form beyond  $\lambda=1$.
These dimers are shown schematically with dashed rectangles
for the two phases in figure (\ref{fig:Schematic-SSH}). 
For a detailed introduction of this model see \cite{su1979solitons, asboth2016short}. 

Here we first show how an \OM array can be tuned 
to mimic the SSH model. Then we use the dynamical tunability of 
the \OM system to change the order parameter and
 emulate the topological phase transition in the SSH model and 
drive the system out of equilibrium.

It is important to note that here  
a modified SSH model is being simulated, namely a bosonic 
SSH model instead of the fermionic one. However, the phase transition 
in question only relates to the properties of the single particle 
wave functions, and hence does not depend on whether we are 
dealing with fermions or bosons.

Next we give the design for the simulator and explain the intuition
behind it. We then present a detailed calculation of the effective
Hamiltonian and show that the Hamiltonian of the simulator is compatible
with the SSH model.

\subsection{Proposal for simulator}

\begin{figure}
\begin{centering}
\includegraphics[width=0.9\columnwidth]{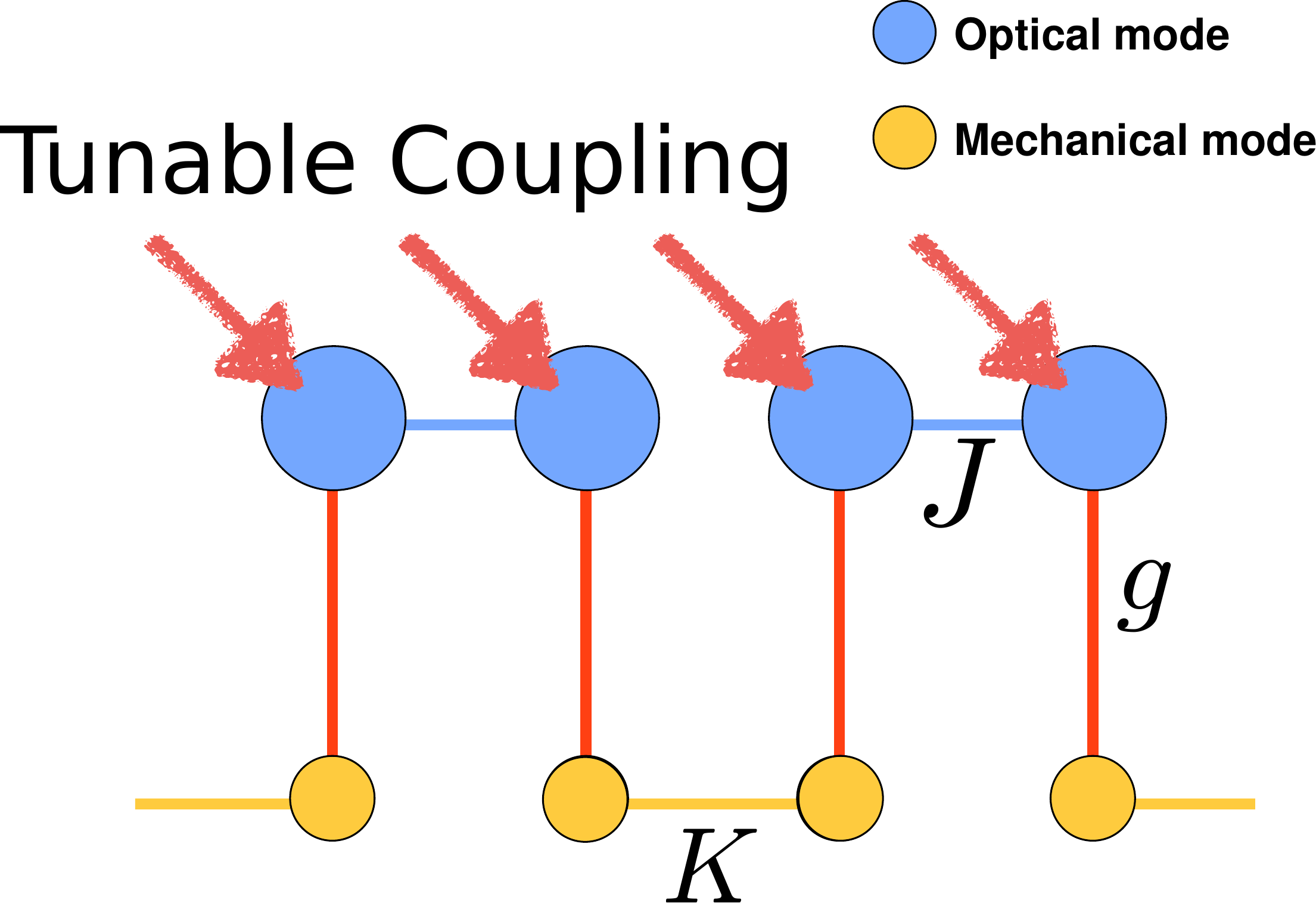} 
\par\end{centering}

\caption{\label{fig:Schematic} Schematic of the 
simulator. Blue/Yellow circles
indicate optical/mechanical modes. 
The mechanical modes in each cell are indirectly coupled
through their coupling to the coupled optical modes. This coupling
can be tuned using an external laser that tunes the optomechanical coupling, $g$. }
\end{figure}

A schematic picture of our design for the \OM simulator
is given in figure (\ref{fig:Schematic}). Such a design can potentially
be implemented in \OM crystals \cite{safavi2014two} and 
electromechanical arrays \cite{lecocq2015quantum}.

We use the mechanical modes as the main modes of the SSH model.
There are two kinds of coupling between the mechanical modes: there
is the direct coupling, $\LatK$,  through the vibrations on the substrate
and the indirect one through the
coupling to the optical modes. The indirect coupling depends on the
direct optical coupling rate, $\LatJ$, and the \OM coupling rate, $g$. 

Next, we calculate the effective Hamiltonian of 
the array in figure (\ref{fig:Schematic}) and find
the indirect coupling with second order perturbation theory.

\subsection{Effective Hamiltonian}

To find the effective Hamiltonian, we focus on one unit cell which
includes two connected \OM nodes (first half of the figure
(\ref{fig:Schematic})). The Hamiltonian of the unit cell is given
by

\begin{align*}
\oprt{H} & =\sum_{i}\left(-\hbar\Delta \oprt{a}_{i}^{\dagger}\oprt{a}_{i}+\hbar\Omega \oprt{b}_{i}^{\dagger}\oprt{b}_{i}\right)\\
 & -\hbar g\sum_{i}\left(\oprt{a}_{i}^{\dagger}\oprt{b}_{i}
 +\oprt{a}_{i}\oprt{b}_{i}^{\dagger}\right)\\
 & +\hbar \LatJ\sum_{{\rm odd}\, i}\left(\oprt{a}_{i}^{\dagger}\oprt{a}_{i+1}
 +\oprt{a}_{i}\oprt{a}_{i+1}^{\dagger}\right)\\
 & +\hbar \LatK\sum_{{\rm even}\, i}\left(\oprt{b}_{i}^{\dagger}\oprt{b}_{i+1}
 +\oprt{b}_{i}\oprt{b}_{i+1}^{\dagger}\right).\label{eq:H_scheme_Raw}
\end{align*}

We  block-diagonalize the subspace corresponding
to the photonic bands and transform the Hamiltonian into
a basis that instead of the original optical modes, is expressed in
terms of the normal modes of the coupled optical cavities. These normal
modes are the symmetric and anti-symmetric superposition of the original
photonic modes, i.e. 
\begin{equation}
\oprt{A}_{\pm}=\frac{\oprt{a}_{1}\pm \oprt{a}_{2}}{\sqrt{2}}.
\end{equation}

For a unit cell, this gives

\begin{align*}
\oprt{H} & =-\hbar\Delta \left(\oprt{A}_{+}^{\dagger}\oprt{A}_{+}
+\oprt{A}_{-}^{\dagger}\oprt{A}_{-}\right) \\
 & +\hbar\Omega \left(\oprt{b}_{1}^{\dagger}\oprt{b}_{1}
 +\oprt{b}_{2}^{\dagger}\oprt{b}_{2}\right)\\
 & - \frac{\hbar g}{\sqrt{2}}
 	\left(\oprt{A}_{+}^{\dagger}\left(\oprt{b}_{1}+\oprt{b}_{2} \right)+
 		  \oprt{A}_{-}^{\dagger}\left(\oprt{b}_{1}-\oprt{b}_{2} \right)+h.c. \right)\\
 & +\hbar J \left(\oprt{A}_{+}^{\dagger}\oprt{A}_{+}-\oprt{A}_{-}^{\dagger}\oprt{A}_{-}\right).
\end{align*}
Note that there are couplings between the mechanical 
modes in the unit cell and the neighbouring cells which 
are not included in the Hamiltonian of the unit cell. 
We will later include them as interaction terms between 
different cells.

The symmetric and anti-symmetric photonic
modes couple to both mechanical modes. 
We use the Bloch matrix of the Hamiltonian above to 
calculate the indirect  coupling 
between the  two mechanical modes with second order perturbation theory, 
which gives  
\begin{align*}
\oprt{H} & =\hbar\hat{\Omega}\sum_{i}\left(\oprt{b}_{i}^{\dagger}\oprt{b}_{i}\right)
\\
 & +\hbar K_{p}\sum_{{\rm odd}\, i}\left(\oprt{b}_{i}^{\dagger}\oprt{b}_{i+1}
 +\oprt{b}_{i}\oprt{b}_{i+1}^{\dagger}\right)\\
 & +\hbar K\sum_{{\rm even}\, i}\left(b_{i}^{\dagger}b_{i+1}+b_{i}b_{i+1}^{\dagger}\right),
\label{eq:H_eff}
\end{align*}
where $K_{p}$  is the effective coupling in the SSH model and is 
\[K_{p}=\frac{2g^{2}J\left(-\Delta^{2}+J^{2}-\Omega^{2}\right)}{(-\Delta+J-\Omega)(\Delta+J-\Omega)(-\Delta+J+\Omega)(\Delta+J+\Omega)}
\]

Note that this can be tuned with $g$. 
Using the parameters that we used for the first part, the
couplings in the SSH model can be estimated as 
$K=3 MHz, 
K_{p}=10 MHz
$.

\begin{figure}
\begin{centering}
\includegraphics[trim={0cm 0cm 0cm 0cm},clip ,width=\linewidth]{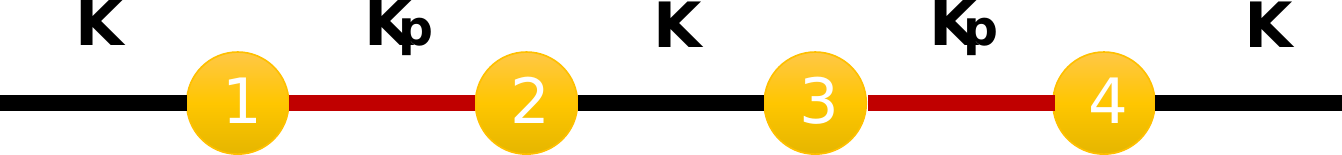}	 
\par\end{centering}

\caption{\label{fig:Heff} Effective model 
after applying second-order perturbation theory to the 
optomechanical array of figure (\ref{fig:Schematic}).  }
\end{figure}

The coupling $K_{p}$ depends on the \OM coupling $g$.
The above numerical estimate for $K_{p}$ is the maximum that can be achieved
using the parameters that we considered here. 
Reducing the laser power, it can be tuned
to $K_{p}<K$ which changes the phase to the non-topological phase. 

This can be used to explore a wide range of properties in 
this system. For instance, we can start in the topological phase with $\lambda >1$, 
with the system initialized in one of the edge states and then 
abruptly change the Hamiltonian to the non-topological phase 
with $\lambda <1$ and probe the evolution of the edge states.

\begin{figure}
\begin{centering}
\includegraphics[trim={0cm 0cm 0cm 0cm},clip ,width=\linewidth]{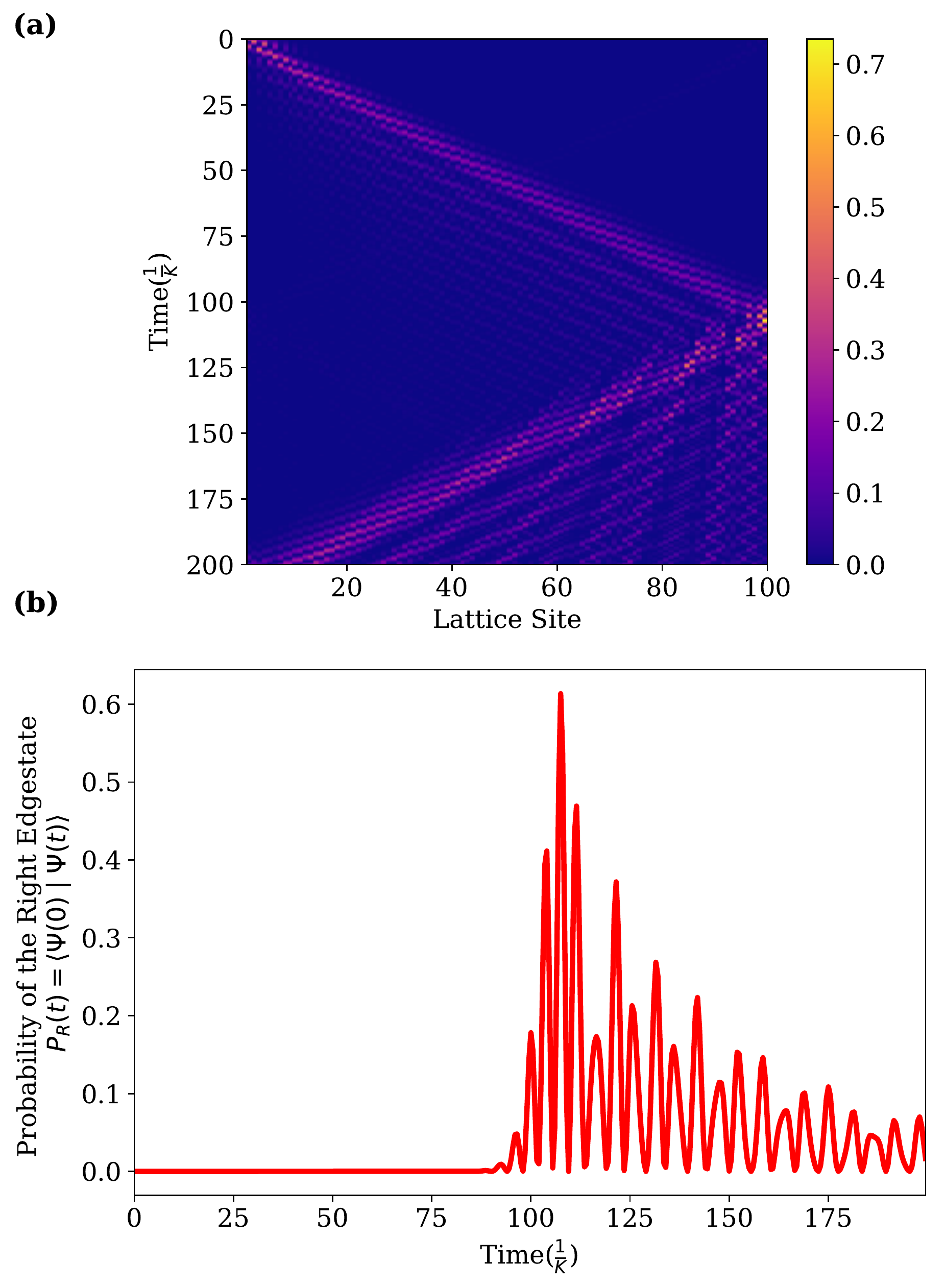}	 
\par\end{centering}

\caption{\label{fig:SSH-edge-space-time} Time evolution of 
an excitation in the SSH model, initiated on the left side 
of the lattice, while abruptly quenching the 
Hamiltonian to the non-topological phase. 
Plot (a) shows the propagation 
of the left edge state in time and space (lattice site).
Plot (b) \label{fig:SSH-edge-time}  shows the probability 
of getting the right edge state in time, i.e. 
$P_{R}(t) = |\langle\Psi(t=0)  \ket{\Psi(t=t)}|^2$. 
At time $t=0$, the excitation is on the leftmost site of the lattice and 
it starts travelling to the right. 
For this simulations, we started with Hamiltonian in the topological phase
and the left edge state for the $\ket{\Psi(t=0)}$. After the quench to  the 
non-topological phase, it takes some time for the excitation to reach 
the right side of the lattice. The plot (b) corresponds to 
the last site of the lattice of (a), indicated in the red box. 
}
\end{figure}

Figure (\ref{fig:SSH-edge-space-time}) shows the 
dynamics of the excitations in this system as 
it evolves through time and space. The excitation 
on the left side of the chain starts to 
propagate to the right after the quench. 
Figure (\ref{fig:SSH-edge-time}-b) shows one 
slice of figure (\ref{fig:SSH-edge-space-time}-a) 
which represents the probability of
observing the excitation on the right side of the lattice after time $t$. 
This probability is negligible at first, and it increases after 
the initially produced excitations have travelled through the whole lattice.

\section{Conclusion}

We have studied  non-equilibrium effects in \OM arrays which can 
 be caused by abrupt changes in the parameters
of the system, induced via the driving laser. 
We have analyzed the resulting excitations and we have shown that the number of such excitations follows a 
power law with respect to the quench speed. 

We have also provided a proposal for exploiting the dynamical aspects
of \OM arrays for simulating non-equilibrium dynamics in the 
SSH model, as a simple example of an array with a band structure that has topological properties. 

We have commented on the experimental outlook. Still, to adopt the first results presented here for concrete experimental platforms, some further, more detailed analysis will be needed. For example, the effects of disorder \cite{roque2017anderson} may need careful additional consideration. 

More generally, the present work paves the way towards investigating other aspects of non-equilibrium 
dynamics in \OM arrays with time-dependent band structures. 
Further studies may reveal which other kinds of phenomena 
should be expected and tested in these settings. 
We emphasize that the \OM system we have considered here 
is in the linear regime, with a quadratic Hamiltonian, 
and is not capable of capturing the complexity of
quantum simulation of non-trivial quantum {\em many-body}
systems. Yet, as we have shown, even the linear dynamics displays
a rich set of features. In the near-term future, one might also
study the nonlinear classical dynamics in nonequilibrium \OM arrays,
which is perfectly within experimental reach.

We thank Vittorio Peano for fruitful discussions. This work was supported
by the ERC Starting Grant OPTOMECH. This work is supported by the research grant system of Sharif
University of Technology (G960219), the European Union’s Horizon 2020 research and innovation programme under grant agreement No 732894 (FET-Proactive HOT).

\bibliography{Quench}


\setcounter{equation}{0} \setcounter{figure}{0} \setcounter{table}{0}
 \global\long\def\thefigure{S\arabic{figure}}
 \global\long\def\bibnumfmt#1{[S#1]}
 \global\long\def\citenumfont#1{S#1}

\pagebreak
\widetext
\begin{center}
\textbf{\large Supplemental Materials:\\ Quench Dynamics in 1D Optomechanical Arrays}
\end{center}
\setcounter{equation}{0}
\setcounter{figure}{0}
\setcounter{table}{0}
\setcounter{page}{1}
\makeatletter
\renewcommand{\theequation}{S\arabic{equation}}
\renewcommand{\thefigure}{S\arabic{figure}}
\renewcommand{\bibnumfmt}[1]{[S#1]}
\renewcommand{\citenumfont}[1]{S#1}

%
%



\subsection*{Normal modes}
Here we give an expression for the normal modes of the Hamiltonian. 

The normal modes are given by the eigenvectors of the $\blochHk$. 
To find the eigenvectors, it helps to rewrite it as 
\begin{equation}
\blochHk = \frac{\Omega(k)-\Delta(k)}{2} \mathbb{{1}} 
-\frac{\Omega(k)+\Delta(k)}{2} \sigma_z +
g \sigma_x.
\end{equation}
Here $\sigma_i$ are the Pauli operators. 
The first term does not affect the eigenvectors. So the eigenvectors
are the eigenvectors of a rotated Pauli operator in the $x-z$ plane. 
With a simple rotation, we can transform the eigenvectors of $\sigma_z$ 
to the eigenvectors of the rotated Pauli operator. For simplicity, 
we define $2\delta(k) =\Omega(k)+\Delta(k) $. With some
simple algebra we get to
\begin{equation}
\vec{\lambda}_{\pm}\left(k\right)=\frac{1}{z_{\pm}}  \left(\begin{array}{c}
-g\pm\sqrt{g^{2}+\delta\left(k\right)^{2}}\\
\delta\left(k\right)
\end{array}\right),
\end{equation}
with $z_{\pm}$ the normalization factors. 
Now if we apply the transformation that diagonalizes the $\blochHk$,
we get 
\begin{align*}
\oprt{H}_{OMA}= & \sum_{k}\hbar\left(\begin{array}{cc}
\oprt{a}_{k}^{\dagger} & \oprt{b}_{k}^{\dagger}\end{array}\right)
\left(\begin{array}{cc}
-\Delta\left(k\right) & g\\
g & \Omega\left(k\right)
\end{array}\right)\left(\begin{array}{c}
\oprt{a}_{k}\\
\oprt{b}_{k}
\end{array}\right)\\
= & \sum_{k}\hbar\left(\begin{array}{cc}
\oprt{a}_{k}^{\dagger} & \oprt{b}_{k}^{\dagger}\end{array}\right)R_{k}^{\dagger}\left(\begin{array}{cc}
\omega_{A}\left(k\right) & 0\\
o & \omega_{B}\left(k\right)
\end{array}\right)R_{k}\left(\begin{array}{c}
\oprt{a}_{k}\\
\oprt{b}_{k}
\end{array}\right)\\
= & \sum_{k}\hbar\left(\begin{array}{cc}
\oprt{A}_{k}^{\dagger} & \oprt{B}_{k}^{\dagger}\end{array}\right)\left(\begin{array}{cc}
\omega_{A}\left(k\right) & 0\\
o & \omega_{B}\left(k\right)
\end{array}\right)\left(\begin{array}{c}
\oprt{A}_{k}\\
\oprt{B}_{k}
\end{array}\right)\\
= & \sum_{k}\hbar\left(\omega_{A}\left(k\right)\oprt{A}_{k}^{\dagger}\oprt{A}_{k}
+\omega_{B}\left(k\right)\oprt{B}_{k}^{\dagger}\oprt{B}_{k}\right).
\end{align*}
Here $\left\{ \omega_{A}\left(k\right),\omega_{B}\left(k\right)\right\} $
are the eigenvalues of the $\blochHk$ and $R_k$ is the matrix
that diagonalizes it. You can see that it gives the transformation 
in Eq. (4) in the main text.

\subsection*{Initial state population}

We need to find the equilibrium population of the normal modes. 
For $g=0$
or too far off resonance, the normal modes are the same as the 
original modes, however, as we approach the avoided crossing points, 
the modes hybridize. 

Before we get to the calculation of the equilibrium population of
the normal modes, it helps to review the same calculation for the simple
case of an isolated mechanical mode. The equation of motion 
for a single mechanical resonator is

\[
\dot{\oprt{b}}\left(t\right)=\left(-i\Omega-\Gamma/2\right)\oprt{b}\left(t\right)+\sqrt{\Gamma}\oprt{b}_{in}\left(t\right),
\]
where $\oprt{b}_{in}$ represents the annihilation operator of the mechanical bath modes. 
This is a simple differential equation which gives 
\[
\oprt{b}\left(t\right)=e^{-i\Omega t-\Gamma t/2}\left(\oprt{b}_{0}\left(t\right)+\sqrt{\Gamma}\int_{0}^{t}dt'e^{i\Omega t'+\Gamma t'/2}\oprt{b}_{in}\left(t\right)\right)
\]

We are interested in $n^{m}=\langle \oprt{b}^{\dagger}\left(t\right)\oprt{b}\left(t\right)\rangle$
which is
\[
n^{m}=\langle \oprt{b}^{\dagger}\left(t\right)\oprt{b}\left(t\right)\rangle
=e^{-\Gamma t}\left(\langle \oprt{b}_{0}^{\dagger}\left(t\right)\oprt{b}_{0}\left(t\right)\rangle
+\Gamma\int_{0}^{t}\int_{0}^{t}dt'dt''e^{\Gamma(t'+t'')/2}e^{\Omega(t'-t'')}\langle \oprt{b}_{in}^{\dagger}\left(t'\right)\oprt{b}_{in}\left(t''\right)\rangle\right).
\]

We make the Markov approximation for the bath which implies that $\langle \oprt{b}_{in}^{\dagger}\left(t'\right)\oprt{b}_{in}
\left(t''\right)\rangle=\delta\left(t'-t''\right)n_{th}^{m}$.
This approximation simplifies the calculation and gives

\begin{align*}
n^{m} & =e^{-\Gamma t}\left(n_{0}^{m}+n_{th}^{m}\Gamma\int_{0}^{t}dt'e^{\Gamma t'}\right)\\
 & =e^{-\Gamma t}\left(n_{0}^{m}+n_{th}^{m}\Gamma\frac{\left(e^{\Gamma t}-1\right)}{\Gamma}\right)\\
 & =e^{-\Gamma t}\left(n_{0}^{m}-n_{th}^{m}\right)+n_{th}^{m}.
\end{align*}

For the stationary state, $t \rightarrow \infty$, 
we get $n^{m}\rightarrow n_{th}^{m}.$

Despite the simplicity, this calculation is the main tool we need to
find the population of the normal modes in the stationary state.  

Consider the equations of motion

\begin{equation}
\dot{\oprt{X}}=M\cdot\oprt{X}+\oprt{\xi}\left(t\right), \label{eq:Schm-EqofMotion}
\end{equation}
where $\oprt{X}=\left(\begin{array}{c}
\oprt{a}\left(t\right)\\
\oprt{b}\left(t\right)
\end{array}\right),\,\xi=\left(\begin{array}{c}
\sqrt{\kappa}\oprt{a}_{in}\left(t\right)\\
\sqrt{\Gamma}\oprt{b}_{in}\left(t\right)
\end{array}\right)$ and 
\begin{equation}
M=\left(\begin{array}{cc}
i\Delta-\frac{k}{2} & ig\\
ig & -i\Omega-\frac{\Gamma}{2}
\end{array}\right).
\end{equation}

Here we are ignoring the amplification terms 
in the Hamiltonian.

For the normal modes, we diagonalize $M$ 
without the dissipation terms. 

This transforms the eq.(\ref{eq:Schm-EqofMotion}) to 
\[
d\left(\begin{array}{c}
\hat{A}\left(t\right)\\
\hat{B}\left(t\right)
\end{array}\right)/dt=\left(\begin{array}{cc}
-i\omega_{A}+\frac{\kappa_{A}}{2} & 0\\
0 & -i\omega_{B}+\frac{\kappa_{B}}{2}
\end{array}\right)\cdot\left(\begin{array}{c}
\hat{A}\left(t\right)\\
\hat{B}\left(t\right)
\end{array}\right)+\left(\begin{array}{c}
\oprt{A}_{in}\left(t\right)\\
\oprt{B}_{in}\left(t\right)
\end{array}\right).
\]
$\{\omega_A , \omega_B \}$ are the frequencies of the normal modes
and $\{\kappa_A , \kappa_B \}$ are the dissipation corresponding to these
modes.  Also $\{\oprt{A}_{in}\left(t\right) , \oprt{B}_{in}\left(t\right) \}$ 
are linear superpositions of $\oprt{a}_{in}\left(t\right), \oprt{b}_{in}\left(t\right)$. 
Note that $\{\kappa_A , \kappa_B \}$ can be calculated as the first order perturbation
to the $M$ without dissipation. More specifically we can take 
\begin{equation}
M=\left(\begin{array}{cc}
i\Delta & ig\\
ig & -i\Omega
\end{array}\right)
+\left(\begin{array}{cc}
-\frac{\kappa}{2} & 0\\
0 & -\frac{\Gamma}{2}
\end{array}\right).
\end{equation}

Now if we focus on the population of the 
normal modes $\hat{A}$ and $\hat{B}$,
it would be the same   calculation that we did for an isolated
mode, except for the fact that $\oprt{A}_{in}$ and $\oprt{B}_{in}$ are now affected
by both the optical and mechanical baths. Repeating the calculations above, 
we get

\begin{equation}
\langle\hat{A}^{\dagger}\left(t\right)\hat{A}\left(t\right)\rangle=e^{-\kappa_{A}t}\left(\langle\hat{A}_{g\left(0 \right)}^{\dagger}\left(t\right)\hat{A}_{g\left(0 \right)}\left(t\right)\rangle+\int_{0}^{t}\int_{0}^{t}dt'dt''e^{\kappa_{A}(t'+t'')/2}e^{\omega_{A}(t'-t'')}\langle \oprt{A}_{in}^{\dagger}\left(t'\right)\oprt{A}_{in}\left(t''\right)\rangle\right)
\end{equation}

Now this requires the calculation of $\oprt{A}_{in}(t)$ and $\oprt{B}_{in}(t)$ which
are given by the transformation $R$. In general
\[
R=\left(\begin{array}{cc}
\epsilon & \gamma\\
\mu & \nu
\end{array}\right),\,\textrm{where }\epsilon^{2}+\gamma^{2}=1\,\textrm{and \ensuremath{\left|\epsilon\right|}=\ensuremath{\left|\nu\right|} and \ensuremath{\left|\gamma\right|}=\ensuremath{\left|\mu\right|}.}
\]
This transforms the modes as $\oprt{A}_{in}\left(t\right)=\epsilon\sqrt{\kappa_{0}}\oprt{a}_{in}
\left(t\right)+\gamma\sqrt{\Gamma}\oprt{b}_{in}\left(t\right)$
and $\oprt{B}_{in}\left(t\right)=\mu\sqrt{\kappa_{0}}\oprt{a}_{in}
\left(t\right)+\nu\sqrt{\Gamma}\oprt{b}_{in}\left(t\right)$ but more importantly, 
\begin{align*}
\langle \oprt{A}_{in}^{\dagger}\left(t\right)\oprt{A}_{in}
\left(t'\right)\rangle= & \epsilon^{2}\kappa_{0}\langle \oprt{a}_{in}^{\dagger}\left(t\right)\oprt{a}_{in}
\left(t'\right)\rangle+\gamma^{2}\Gamma\langle \oprt{b}_{in}^{\dagger}\left(t\right)\oprt{b}_{in}\left(t'\right)\rangle\\
\langle \oprt{B}_{in}^{\dagger}\left(t\right)\oprt{B}_{in}\left(t'\right)\rangle= & \mu^{2}\kappa_{0}\langle \oprt{a}_{in}^{\dagger}\left(t\right)\oprt{a}_{in}
\left(t'\right)\rangle+\nu^{2}\Gamma\langle \oprt{b}_{in}^{\dagger}\left(t\right)\oprt{b}_{in}\left(t'\right)\rangle.
\end{align*}
Now recall that we are using the Markov approximation and 
since the optical bath is at zero temperature, 
we get
\begin{align*}
\langle\hat{A}^{\dagger}\left(t\right)\hat{A}\left(t\right)\rangle= & e^{-\kappa_{A}t}\left(\langle\hat{A}_{g\left(0 \right)}^{\dagger}\left(t\right)\hat{A}_{g\left(0 \right)}\left(t\right)\rangle+\int_{0}^{t}\int_{0}^{t}dt'dt''e^{\kappa_{A}(t'+t'')/2}e^{\omega_{A}(t'-t'')}\mid\gamma\mid^{2}\Gamma\delta\left(t'-t''\right)n_{th}^{m}\right)\\
= & e^{-\kappa_{A}t}\left(\langle\hat{A}_{g\left(0 \right)}^{\dagger}\left(t\right)\hat{A}_{g\left(0 \right)}\left(t\right)\rangle+\mid\gamma\mid^{2}\Gamma n_{th}^{m}\int_{g0}^{t}dt'e^{\kappa_{A}t'}\right)\\
= & e^{-\kappa_{A}t}\left(\langle\hat{A}_{g\left(0 \right)}^{\dagger}\left(t\right)\hat{A}_{g\left(0 \right)}\left(t\right)\rangle+\mid\gamma\mid^{2}\Gamma n_{th}^{m}\left(\frac{e^{\kappa_{A}t}-1}{\kappa_{a}}\right)\right)\\
\left(\underset{t\rightarrow\infty}{\lim}\right)= & \frac{\mid\gamma\mid^{2}\Gamma n_{th}^{m}}{\kappa_{A}}
\end{align*}

With calculation of $\kappa_A$ and $\kappa_B$, we get 
\begin{align*}
n_{th}^{m}\left(\hat{A}\right)= & \frac{ p \Gamma n^M_{th}}{(1-p)\kappa + p \Gamma}\\
n_{th}^{m}\left(\hat{B}\right)= & \frac{\left(1-p\right) \Gamma n^M_{th} }{p \kappa + (1-p)\Gamma}.
\end{align*}
where $p_k$ is given by the projection of the normal mode $\oprt{A}_k$ 
on the original mode $\oprt{a}_k$.

\subsection*{Occupation of the modes and their evolution} 
We are interested in 
\begin{align*}
 & \bra{\psi(t)} \Oprtk{A}{g\left(t\right)}^{\dagger} \Oprtk{A}{g\left(t\right)} \ket{\psi(t)}\\
=  & \bra{\psi(0)} U(t)^{\dagger} \Oprtk{A}{g\left(t\right)}^{\dagger} \Oprtk{A}{g\left(t\right)} U(t)\ket{\psi(0)}\\
=  & \bra{\psi(0)}\underset{\widetilde{A}^{\dagger}\left(t\right)}{\underbrace{U(t)^{\dagger}\Oprtk{A^{\dagger}}{g\left(t\right)}U(t)}}\underset{\widetilde{A}\left(t\right)}{\underbrace{U(t)^{\dagger}\Oprtk{A}{g\left(t\right)}U(t)}}\ket{\psi(0)}
\end{align*}
Similarly, we can define $\widetilde{B}\left(t\right)$.
Note that we drop the subscript $k$ for simplicity. 
Our goal is to express 
$\left\{ \widetilde{A}(t),\widetilde{B}(t)\right\} $ 
in terms of the 
$\left\{ \oprt{A}_{g\left(0\right)},\oprt{B}_{g\left(0\right)}\right\}$.
This is because we already calculated the occupation number of the initial normal modes, 
i.e. for $g(0)$. Also, for the initial mode, the cross expectation values like  
$\left\langle \oprt{A}_{g\left(0\right)}^{\dagger}\oprt{B}_{g\left(0\right)}\right\rangle $
 vanish.

To this end, we use the Eq. (\ref{A(t,t)}) in the main text. 
Just note that we first express 
$\left\{ \widetilde{A}(t),\widetilde{B}(t)\right\} $ 
in terms of $\left\{ \oprt{a} , \oprt{b} \right\} $  and then 
we inverse the equation to express it in terms of 
$\left\{ \oprt{A}_{g\left(0\right)},\oprt{B}_{g\left(0\right)}\right\}$. This gives
\[
\left(\begin{array}{c}
\widetilde{A}\left(t\right)\\
\widetilde{B}\left(t\right)
\end{array}\right)=R\left(g\left(t\right)\right)S\left(t\right)R^{-1}\left(g\left(0\right)\right)\left(\begin{array}{c}
\oprt{A}_{g\left(0\right)}\\
\oprt{B}_{g\left(0\right)}
\end{array}\right)
\]

This gives $\widetilde{A} = c_1 \oprt{A} + c_2 \oprt{B}$, where $c_1$ and $c_2$ are two coefficient extracted from 
equation above and for simplicity we dropped the time and the subscripts. Now the population 
of these new mode would be
\begin{align*}
 & \bra{\psi(0)}\widetilde{A}^{\dagger}\left(t\right)\widetilde{A}
 \left(t\right)\ket{\psi(0)}\\
= & \left|c_{1}\right|^{2}n\left(\oprt{A}\left(g\left(0\right)\right)\right)+\left|c_{2}\right|^{2}n\left(\oprt{B}\left(g\left(0\right)\right)\right)
\end{align*}
where  $n\left(\oprt{A}_{g\left(0\right)}\right)$ and $n\left(\oprt{B}_{g\left(0\right)}\right)$
can be calculated from the previous section. 

\end{document}